\documentclass[twocolumn,prb,showpacs,superscriptaddress]{revtex4}
\usepackage{graphicx}
\usepackage{amsmath}    
\def\w{\omega}
\def\wp{\omega^\prime}
\def\wc{{\omega_{\rm C}}}
\def\>{\rangle}
\def\<{\langle}
\def\H{\hat{H}}
\def\P{\hat{P}_{\rm occ}}
\def\E{\varepsilon}
\def\vp{{v^\prime}}
\def\q{{\bf q}}
\def\s{\sigma}
\def\k{{\bf k}}

\def\G{{\bf G}}
\def\Gp{{\bf G^\prime}}
\def\rt{\tilde{r}}
\def\pt{\tilde{p}}
\def\r{{\bf r}}
\def\rp{{\bf r^\prime}}
\def\rpp{{\bf r^{\prime\prime}}}
\def\rppp{{\bf r^{\prime\prime\prime}}}
\def\mo{$\overline{1}$}
\def\mt{$\overline{2}$}

\begin{document}

\title{GW method with the self-consistent Sternheimer equation}

\author{Feliciano Giustino}
\email{feliciano.giustino@materials.ox.ac.uk}
\affiliation{Department of Materials, University of Oxford, Parks Road, Oxford OX1 3PH, United Kingdom}
\affiliation{Department of Physics, University of California at Berkeley, 
Berkeley, California 94720, USA,
and Materials Sciences Division, Lawrence Berkeley National Laboratory, 
Berkeley, California 94720, USA}
\author{Marvin L. Cohen}
\author{Steven G. Louie}
\affiliation{Department of Physics, University of California at Berkeley, 
Berkeley, California 94720, USA,
and Materials Sciences Division, Lawrence Berkeley National Laboratory, 
Berkeley, California 94720, USA}
\date{\today}

\begin{abstract}
We propose a novel approach to quasiparticle GW calculations which 
does not require the computation of unoccupied electronic states. 
In our approach the screened Coulomb interaction
is evaluated by solving self-consistent linear-response Sternheimer equations,
and the noninteracting Green's function is evaluated by solving inhomogeneous 
linear systems. The frequency-dependence of the screened Coulomb interaction 
is explicitly taken into account. In order to avoid the singularities of the 
screened Coulomb interaction the calculations are performed
along the imaginary axis, and the results are analytically continued 
to the real axis through Pad\'e approximants. As a proof of concept we implemented 
the proposed methodology within the empirical pseudopotential formalism and 
we validated our implementation using silicon as a test case. We examine 
the advantages and limitations of our method and describe promising future directions.
\end{abstract}

 % Excited states: methodology

\maketitle

\section{Introduction}

During the past two and a half decades the $GW$ method\cite{hedin1,hl86}
for the study of electron quasiparticle excitations
has had a number of successes and witnessed significant growth of interest
within the computational electronic structure community.
The $GW$ method is currently being used for predicting electron quasi-particle
excitation spectra as well as optical spectra in a variety of materials
ranging from bulk solids to nanostructures and organic systems. The
$GW$ method is also of widespread use as a starting point for Bethe-Salpeter calculations
of two-particle neutral excitations.\cite{onida,bse1,bse2,bse3,rolfing,reining-review}
Current implementations find many diverse applications,
including among others the calculation of the optical response of nanostructures,\cite{catalin} quantum transport in nanoscale
junctions,\cite{rubio} pump-probe spectroscopy,\cite{catalin-lw} angle-resolved photoemission
spectroscopy,\cite{cheolhwan} and strongly-correlated systems.\cite{bruneval-oxide}

Current trends in the development of improved computational approaches for
quasiparticle excitations based on the $GW$ method include the refinement
of the initial guess for the non-interacting Green's function and for the
polarization operator,\cite{rinke,schilfegarde}
the inclusion of approximate vertex corrections or higher-order self-energy
diagrams,\cite{york} and the description of the frequency-dependent
dielectric response beyond the original generalized plasmon-pole approximation.\cite{spacetime,blochl}
Detailed reviews of past and current developments in $GW$ techniques can be
found in Refs.\ \onlinecite{hl,gunnarsson,reining-review,rinke,wilkins}.

The majority of current $GW$ implementations obtain the screened Coulomb interaction
$W$ and the non-interacting Green's function $G$ using a perturbative
expansion over the Kohn-Sham eigenstates (cf. Sec.\ \ref{sec.coulomb} below). 
Such expansion requires the calculation of both occupied and unoccupied electronic states,
as well as their associated optical matrix elements.\cite{hl86} A common bottleneck 
of this approach is that the convergence of the quasi-particle excitation energies
with the number 
of unoccupied states is rather slow.\cite{sohrab} This difficulty is particularly relevant
when calculating the absolute values of the quasiparticle excitation energies.\cite{bruneval-gonze}
Several avenues have been explored so far in order to circumvent this bottleneck 
and to perform $GW$ calculations by employing only occupied electronic states,\cite{reining-sternheimer,umari1,umari2,gygi}
or a small number of unoccupied states.\cite{bruneval-gonze}

The main aim of the present work is to demonstrate the feasibility of $GW$ calculations
entirely based on occupied states only.\cite{baroni.rmp}
In practice we adopt the principles of density-functional perturbation theory (DFPT) 
to determine (i) the frequency-dependent screened Coulomb interaction
by directly solving self-consistent linear response Sternheimer equations,
and (ii) the non-interacting Green's function by solving
inhomogeneous linear systems. The main advantage of the proposed method 
is that it does not require the computation of unoccupied electronic states.
In addition, we demonstrate the possibility of fast evaluations of 
the frequency dependence of the screened Coulomb interaction
based on {\it multishift} linear-system solvers.\cite{frommer}
As a proof of concept we have implemented our method within
a planewaves empirical pseudopotential scheme,\cite{cohen_berg} 
and validated it by comparing with previous work for the prototypical test case of silicon.

The use of the Sternheimer equation for calculating the polarizability
in the random-phase approximation (RPA) or the inverse dielectric matrix 
has already been discussed in Refs.\ \onlinecite{fleszar-resta} and \onlinecite{kunc-tosatti},
respectively, 
within the framework of a non-perturbative supercell approach. After the introduction 
of DFPT in the context of lattice-dynamical calculations,\cite{giannozzi} 
the authors of Ref.\ \onlinecite{reining-sternheimer} proposed the
use of the non self-consistent Sternheimer method for the calculation
of the dielectric matrix. The elimination of unoccupied electronic states
in the evaluation of the screened Coulomb interaction has also been 
proposed recently within the framework of a Wannier-like representation
of the polarization propagator and the Lanczos recursion method.\cite{umari1,umari2}

This manuscript is organized as follows. In Sec.\ \ref{sec.theory} we describe
how the self-consistent Sternheimer formalism can be adapted to perform $GW$ calculations.
In particular, we outline the procedure to obtain the screened Coulomb
interaction in Sec.\ \ref{sec.coulomb}, the non-interacting Green's function
in Sec.\ \ref{sec.green}, and the $GW$ self-energy in Sec.\ \ref{sec.sigma}.
In Sec.~\ref{sec.theory.g} we specialize to a planewave basis set representation
and derive the key equations for the case of Bloch electrons.
Sections \ref{sec.coulomb.g}, \ref{sec.green.g}, \ref{sec.sigma.g}
parallel the corresponding sections in the general theory part, respectively.
In Sec.\ \ref{sec.scaling} we critically analyze the advantages 
and limitations of the present approach with an emphasis on the
scaling of the calculations with system size. 
In Sec.\ \ref{sec.results} we report the results of our test calculations
for silicon and compare with previous calculations in the literature. Specifically,
we presents results for the direct and inverse dielectric matrix (Sec.\ \ref{sec.5a}),
for the analytic continuation of the dielectric matrix using Pad\'e approximants
(Sec.\ \ref{sec.conv}), for the self-energy (Sec.\ \ref{sec.5c}), 
and for the spectral function (Sec.\ \ref{sec.5d}).
In Sec.~\ref{sec.conclusions} we discuss possible
future developments of our method and discuss our conclusions.
The Appendices provide technical details on some numerical algorithms adopted
in this work, in particular the preconditioned complex biconjugate gradient method (Appendix~\ref{app.cbcg}),
the analysis of the conditioning of the Sternheimer equations (Appendix~\ref{app.condition}),
the analytic continuation using Pad\'e approximants (Appendix \ref{app.pade}),
and the use of multishift methods for the simultaneous calculation of 
the polarization at multiple frequencies (Appendix \ref{app.multishift}).

\section{General theory}\label{sec.theory}

\subsection{Screened Coulomb interaction}\label{sec.coulomb}

In this section we describe how to exploit the Sternheimer scheme within density-functional
perturbation theory in order to calculate the screened Coulomb interaction
$W(\r,\rp;\w)$ (where $\r$, $\rp$ are the space variables and $\w$ is the
excitation frequency). While the use of the Sternheimer approach in DFPT 
was originally developed bearing in mind the Kohn-Sham effective Hamiltonian, 
we note that the present procedure applies without restrictions also to 
post-DFT methods such as the LDA+U method,\cite{anisimov} hybrid 
functionals,\cite{becke} and exact exchange.\cite{exx}
We assume Rydberg atomic units throughout this manuscript. 
The Hedin's equation which defines the screened Coulomb 
interaction reads:\cite{hl}
  \begin{eqnarray}\label{eq.w}
  W(\r,\rp;\w) & = & v(\r,\rp) + \int d\rpp \,W(\r,\rpp;\w)  \nonumber \\
   & \times & \int d\rppp P(\rpp,\rppp;\w) v(\rppp,\rp),
  \end{eqnarray}
where $v(\r,\rp)=e^2/|\r-\rp|$ denotes the bare Coulomb interaction and 
$P(\r,\rp;\w)$ the irreducible polarization propagator. 
As Eq.\ (\ref{eq.w}) is a self-consistent Dyson equation for the
screened Coulomb interaction, it should be possible to solve it
recursively in the spirit of density-functional perturbation theory.
For simplicity, we here specialize to the case of the random-phase approximation (RPA)
for the polarization propagator. 
The generalization of this procedure to include exchange and correlation
effects can be performed without difficulties (cf.\ Sec.~\ref{sec.vertex}).
Within the random-phase approximation the polarization propagator can be written as:\cite{hl}
  \begin{equation}\label{eq.p}
  P(\r,\rp;\w) = 2\sum_{n m} \frac{f_n-f_m}{\E_n-\E_m-\w} 
  \psi_n(\r)\psi_m^\star(\r)  \psi_n^\star(\rp)\psi_m(\rp),
  \end{equation}
where $\psi_n(\r)$ indicates an electronic eigenstate of the
single-particle Hamiltonian 
with energy eigenvalue $\E_n$ and occupation number $f_n$. 
In the following we assume that the $\psi_n(\r)$ are Kohn-Sham
eigenstates for definiteness.
In Eq.\ (\ref{eq.p}) the summation indices $m$ and $n$ run over
both occupied and unoccupied electronic states, and the factor of 2 accounts for the
spin degeneracy.\cite{hl}
Although the expression for the RPA polarization Eq.~(\ref{eq.p})
has been derived for real frequencies in Ref.\ \onlinecite{hl}, it
is possible to continue the polarization throughout the complex 
plane by using Eq.~(\ref{eq.p}) as a definition outside of the real axis.

Our goal is to rewrite Eqs.\ (\ref{eq.w}) and (\ref{eq.p})
by avoiding explicit summations over the unoccupied electronic states.
For this purpose it is convenient to regard the screened Coulomb interaction
$W(\r,\rp;\w)$ as a function of the 
second space variable $\rp$, whilst the first space variable 
$\r$ and the frequency $\w$ are kept as parameters: $\Delta V_{[\r,\w]}(\rp) = W(\r,\rp;\w)$.
If the system under consideration is subject to the perturbation
$\Delta V_{[\r,\w]}(\rp)$, then within the RPA the first-order variation 
of the single-particle density matrix $\Delta n_{[\r,\w]}$ reads
  \begin{equation}\label{eq.deltan}
  \Delta n_{[\r,\w]} = 2\sum_{v\s} \psi_v^\star  \Delta \psi^\s_{v[\r,\w]}.
  \end{equation}
In Eq.\ (\ref{eq.deltan}) the index $v$ stands for ``valence'' and runs
over the occupied electronic states only, the factor of 2 is for the spin degeneracy, 
and the superscript $\sigma=\pm$ refer to the positive and negative
frequency components of the induced charge.
The first-order variations of the occupied wavefunctions $\Delta \psi^\s_{v[\r,\w]}$
can be determined by solving the following two Sternheimer equations:
  \begin{equation}\label{eq.linsys.1}
  (\H-\E_v\pm\w) \Delta \psi^\pm_{v[\r,\w]}  = -(1-\P)  \Delta V_{[\r,\w]} \psi_v, 
  \end{equation}
where $\H$ is the effective single-particle Hamiltonian and 
$\P=\sum_v |\psi_v\rangle\langle\psi_v|$ is the projector
on the occupied states manifold. 
In the particular case of vanishing frequency ($\w=0$)
the $\sigma=\pm$ variations of the wavefunctions do coincide,
and the standard DFPT equations are recovered.
The screening Hartree potential associated with the induced charge 
$\Delta n_{[\r,\w]}$ is calculated as usual through
  \begin{equation}\label{eq.dhartree}
  \Delta V^{\rm H}_{[\r,\w]}(\rp) = \int d\rpp \Delta n_{[\r,\w]} (\rpp) \, v(\rpp,\rp),
  \end{equation}
and finally the screened Coulomb interaction in the RPA is obtained as 
  \begin{equation}\label{eq.w.dfpt}
  W(\r,\rp;\w) = \Delta V_{[\r,\w]}(\rp) = v(\r,\rp) + \Delta V^{\rm H}_{[\r,\w]}(\rp).
  \end{equation}

It is tedious but otherwise straightforward to verify that 
Eqs.\ (\ref{eq.deltan})-(\ref{eq.w.dfpt}) are {\it equivalent} to the 
original Eqs.\ (\ref{eq.w})-(\ref{eq.p}).
The only assumptions made in our derivation are that 
time-reversal symmetry applies,
and that the system under consideration has a finite 
energy gap for electronic excitations. 
The assumption of time-reversal symmetry is not essential and is mainly used to
obtain a compact expression for the $\sigma=\pm$ wavefunction perturbations.
The assumption of finite energy gap can be relaxed by using the extension
of DFPT to metallic systems developed in Ref.\ \onlinecite{degironcoli}. 

There is a simple and intuitive physical meaning associated with 
the calculation scheme outlined above. To see this
we consider an external {\it test charge} introduced in the system
at the point $\r$. This charge generates a bare Coulomb potential $v(\r,\rp)$,
and the system responds to such perturbation by generating the induced charge
$\Delta n_{[\r,\w]}(\rp)$ and the associated 
screening potential $\Delta V^{\rm H}_{[\r,\w]}(\rp)$.
The sum of the external perturbation $v(\r,\rp)$ and the screening
potential $\Delta V^{\rm H}_{[\r,\w]}(\rp)$ yields the screened Coulomb
interaction $W(\r,\rp;\w)$ at the point $\rp$ within the RPA.

The linear systems in Eq.\ (\ref{eq.linsys.1}) must be solved self-consistently.
For this purpose we begin by initializing the screened Coulomb
interaction $W$ using the bare interaction $v$. 
We then calculate the linear variations of the wavefunctions $\Delta \psi_v^\sigma$.
Using the calculated linear variations we update the induced charge density $\Delta n$
and the associated screening potential $\Delta V^{\rm H}$. This allow us to
generate an improved estimate of the screened Coulomb interaction $W$.
We cycle through these steps until convergence of the screened Coulomb
interaction is achieved.
Equations (\ref{eq.deltan}),(\ref{eq.linsys.1}) can be regarded as
the generalization of the self-consistent Sternheimer equations used for lattice-dynamical 
calculations\cite{baroni.rmp} to finite-frequency test-charge perturbations.

In practical calculations we solve Eq.\ (\ref{eq.linsys.1}) along the imaginary
frequency axis in order to avoid the null eigenvalues of the operator
$\H-\E_v\pm\w$, and then we perform the analytic continuation of 
the screened Coulomb interaction to real frequencies (cf.\ Appendixes \ref{app.cbcg}-\ref{app.pade}).
In the special case of $\w=0$ it is convenient to modify the linear operator on the left-hand side of
Eq.\ (\ref{eq.linsys.1}) by adding the projector on the occupied states manifold $\P$:
 \begin{equation} \label{eq.linsys.1b}
 (\H-\E_v+\alpha\P) \Delta \psi_{v[\r,0]}  = -(1-\P)  \Delta V_{[\r,0]} \psi_v,
  \end{equation}
with $\alpha$ set to twice the occupied bandwith. This extra term does not affect
the solutions $\Delta \psi_{v[\r,0]}$ which are linear combinations
of unoccupied electronic states. At the same time, the extra term
has the effect of shifting away from zero the null eigenvalues 
of the linear operator $\H-\E_v$ thereby making it non singular.
This strategy is common practice in DFPT implementations,\cite{baroni.rmp,espresso} and
is discussed in greater detail in Appendix \ref{app.condition}.

\subsubsection{Vertex correction}\label{sec.vertex}

Within the scheme outlined here it is rather straightforward to introduce
an approximate vertex correction to the $GW$ self-energy along the lines 
of Refs.\ \onlinecite{hl86,reining94}. This correction results from setting
the self-energy in the first iteration of Hedin's equations to the DFT
exchange-correlation (XC) potential, $\Sigma_0(\r,\rp;\w) = \delta(\r,\rp) V_{\rm xc}(\r)$.
Within the present scheme this correction is simply obtained by including
the variation of the exchange-correlation
potential in the self-consistent potential used in Eq.\ (\ref{eq.linsys.1}):
  \begin{equation}\label{eq.linsys.vertex}
  (\H-\E_v\pm\w) \Delta \psi^\pm_{v[\r,\w]}\!  = \!-(1-\P)\!  \Big[\Delta V_{[\r,\w]} + K_{\rm xc}\Delta n_{[\r,\w]}\Big] \! \psi_v,
  \end{equation}
$K_{\rm xc}=\delta V_{\rm xc}/\delta n$ being the functional derivative of the XC
potential with respect to the density. The screened Coulomb interaction is still
to be calculated through Eq.~(\ref{eq.w.dfpt}). This approach has been called 
``$GW+K_{\rm xc}$'' approximation in Ref.\ \onlinecite{reining94} due to the inclusion of the XC contribution
in the screening of the test charge. 
That the inclusion of the XC term in the self-consistent
induced potential leads to the $GW+K_{\rm xc}$ approximation can easily be seen as follows.
We combine Eqs.\ (\ref{eq.deltan}),(\ref{eq.linsys.vertex}) to yield
the induced charge density (we use symbolic operator notation for clarity):
 \begin{equation}
 \Delta n = v\,[1- P (v+K_{\rm xc})]^{-1}P .
 \end{equation}
Then, we substitute this result in the definition of the screened Coulomb interaction
Eqs.\ (\ref{eq.dhartree}),(\ref{eq.w.dfpt}) to find
 \begin{equation}
 W = v \,\{ 1 + v [1-P(v+K_{\rm xc})]^{-1}P \}.
 \end{equation}
The last equation yields precisely the screened Coulomb interaction in the $GW+K_{\rm xc}$ approximation.\cite{hl86,reining94}
The difference between this approach and the
standard $GW$ approximation is that in this case the screening
charge is calculated for an {\it electron}, while in the $GW$-RPA approximation
the screening is calculated for a {\it test charge}.
It is worth pointing out that in standard implementations of DFPT the
XC term is already included in the variation of the self-consistent potential,\cite{baroni.rmp}
therefore the use of the $GW+K_{\rm xc}$ approximation would not require any additional
computational developments if the 
present approach was to be implemented on top of existing DFT software.

\subsubsection{Non self-consistent calculation of the dielectric matrix}\label{sec.diel.nscf}

An alternative approach to the calculation of the screened Coulomb interaction
using the self-consistent Sternheimer equation consists in solving Eq.\ (\ref{eq.linsys.1}) non self-consistently.
For this purpose we can replace the self-consistent perturbation $\Delta V_{[\r,\w]}(\rp)$
in the right-hand side of Eq.\ (\ref{eq.linsys.1}) by the bare Coulomb potential
$v_{[\r]}(\rp)=v(\r,\rp)$ as follows:
  \begin{equation}\label{eq.linsys.1.nscf}
  (\H-\E_v\pm\w) \Delta \psi^{{\rm NS},\pm}_{v[\r,\w]}  = -(1-\P)  v_{[\r]} \psi_v, 
  \end{equation}
and we can solve this Sternheimer equation with the known term on the right-hand side
kept fixed. By constructing the non self-consistent induced charge density $\Delta n^{\rm NS}_{[\r,\w]}$ as in Eq.\ (\ref{eq.deltan})
we then obtain the dielectric matrix $\epsilon(\r,\rp;\w)$:
  \begin{equation}\label{eq.diel}
  \epsilon(\r,\rp;\w) = \delta(\r,\rp) - \Delta n^{\rm NS}_{[\r,\w]}(\rp).
  \end{equation}
It is straightforward to check that this procedure correctly leads to the RPA
dielectric matrix.\cite{hl86-prb}
The difference between this approach and the self-consistent calculation described in
Sec.\ \ref{sec.coulomb} is that here we also need to invert the dielectric matrix obtained
through Eq.\ (\ref{eq.diel}) in order to calculate the screened Coulomb interaction.

This non self-consistent procedure was first proposed in Ref.\ \onlinecite{reining-sternheimer}.
One additional step that we make in the present work is to notice that Eq.\ (\ref{eq.linsys.1.nscf})
constitutes a {\it shifted} linear system, i.e.\ a system where the linear operator
$\H-\E_v\pm\w$ differs from the ``seed'' operator $\H-\E_v$ only by a constant shift $\pm\w I$ 
($I$ being the identity operator). 
In this case we can take
advantage of the {\it multishift} linear system solver of Ref.\ \onlinecite{frommer}
to determine $\Delta \psi^{{\rm NS},\pm}_{v[\r,\w]}$ for every frequency $\w$
at the computational cost of one single calculation for the {\it seed} system $(\H-\E_v) \Delta \psi^{{\rm NS}}_{v[\r,0]}  = -(1-\P)  v_{[\r]} \psi_v$.
This procedure is extremely advantageous as it makes it possible to calculate the entire frequency-dependence 
by performing one single iterative minimization. 
The technical implementation of this procedure is described in Appendix \ref{app.multishift}.

\subsection{Green's function}\label{sec.green}

The calculation of the Green's function can efficiently be performed 
by adopting a strategy similar to the Sternheimer approach
described in Sec.\ \ref{sec.coulomb}.
We introduce the noninteracting Green's function following Ref.\ \onlinecite{hl}:
  \begin{equation}\label{eq.green.1}
  G(\r,\rp;\w) = \sum_n \frac{\psi_n(\r)\psi_n^\star(\rp)}{\w-\E_n-i\eta_n},
  \end{equation}
where the sum extends over occupied as well as unoccupied electronic states.
The real infinitesimal $\eta_n$ is positive ($\eta_n=\eta$) 
for occupied states and negative ($\eta_n=-\eta$) for unoccupied states.\cite{hl,hl86,note.spinfactor}
We now split the sum in Eq.~(\ref{eq.green.1}) into occupied ($v$) and unoccupied ($c$) states:
  \begin{equation}\label{eq.green.2}
  G(\r,\rp;\w) = \sum_v \frac{\psi_v(\r)\psi_v^\star(\rp)}{\w-\E_v-i\eta}
  + \sum_c \frac{\psi_c(\r)\psi_c^\star(\rp)}{\w-\E_c+i\eta},
  \end{equation}
and we add and subtract $\sum_v \psi_v\psi_v^\star/(\w-\E_v+i\eta)$ to obtain:
  \begin{equation}\label{eq.green.split}
  G(\r,\rp;\w) = G^{\rm A}(\r,\rp;\w) + G^{\rm N}(\r,\rp;\w),
  \end{equation}
with
  \begin{eqnarray}\label{eq.green.split.2}
\!\!\!\!\!\!  G^{\rm A}(\r,\rp;\w) & =&  \sum_n \frac{\psi_n^\star(\r)\psi_n(\rp)}{\w-\E_n^-},  \\ 
\!\!\!\!\!\!  G^{\rm N}(\r,\rp;\w)  & = &  2\pi i \sum_v \delta(\w-\E_v) \psi_v^\star(\r)\psi_v(\rp). \label{eq.green.split.3} 
  \end{eqnarray}
In the above derivation we assumed again time-reversal symmetry, we used the Lorentzian representation of the Dirac's delta function 
for small $\eta$ [$\pi\delta(x)=\eta/(x^2+\eta^2)$], and we defined $\E_n^- = \E_n - i\eta$.
The component $G^{\rm A}$ of the Green's function is obviously analytic in the
upper half of the complex energy plane as its poles lie below the real axis.
The non-analytic component $G^{\rm N}$ vanishes
whenever the frequency $\w$ is above the chemical potential. For frequencies $\w$
below the chemical potential, the non-analytic component introduces the poles associated with the occupied 
electronic states. 
The partitioning of Eqs.\ (\ref{eq.green.split.2}),
(\ref{eq.green.split.3}) closely reflects the analytic structure
of the non-interacting Green's function.
A detailed discussion of this aspect can be found in Ref.\ \onlinecite{hl}.
The two components of the Green's function in Eq.\ (\ref{eq.green.split.2}) 
are associated with the Coulomb hole (COH) and the screened exchange (SEX) terms of the
self-energy, 
$\Sigma^{\rm COH} = G^{\rm A} W$  and $\Sigma^{\rm SEX} = G^{\rm N} W$,
respectively.\cite{hl86}

The computation of the non-analytic component $G^{\rm N}$ of the
Green's function in Eq.\ (\ref{eq.green.split.3}) is straightforward 
once the occupied electronic eigenstates have been determined.
In order to calculate the analytic component $G^{\rm A}$ it is convenient to
proceed as in the case of the screened Coulomb interaction,
by regarding $G^{\rm A}(\r,\rp;\w)$ as a parametric
function of the the first space variable and of the frequency:
$G^{\rm A}_{[\r,\w]}(\rp) = G^{\rm A}(\r,\rp;\w)$.
If we apply the operator $\H-\w^+$ to both sides of Eq.\ (\ref{eq.green.split.2}),
with $\w^+=\w+i\eta$,
and we use the completeness relation $\delta_{[\r]}(\rp) = \delta(\r,\rp) = \sum_n \psi_n(\r)\psi_n^\star(\rp)$, then we find immediately:
  \begin{equation}\label{eq.green.3}
  (\H-\w^+) G^{\rm A}_{[\r,\w]} = -\, \delta_{[\r]}.
  \end{equation}
As expected, we can determine the analytic part of the Green's function
by directly solving a linear system. As Eq.\ (\ref{eq.green.3}) 
does not explicitly require unoccupied electronic states, this procedure mimics 
the Sternheimer approach for the screened Coulomb interaction outlined in Sec.~\ref{sec.coulomb},
albeit without the self-consistency requirement.

The procedure described here is especially advantageous because Eq.\ (\ref{eq.green.3})
constitutes a shifted linear system, in the same way as Eq.\ (\ref{eq.linsys.1.nscf}).
Also in this case we exploit the multishift method of Ref.\ \onlinecite{frommer}
to determine $G^{\rm A}_{[\r,\w]}$ for every frequency $\w$
at the computational cost of one single calculation for the seed system $\H G^{\rm A}_{[\r,0]} = -\, \delta_{[\r]}$
(cf.\ Appendix \ref{app.multishift}).

The presence of the infinitesimal $i\eta$ in $\w^+=\w+i\eta$ 
guarantees that the linear operator $\H-\w^+$ in Eq.\ (\ref{eq.green.3.g}) is
never singular.
This operator can nonetheless become ill-conditioned, hence the use of appropriate
preconditioners may become necessary. We discuss this aspect 
in Appendix~\ref{app.condition}.

\subsection{Self-energy}\label{sec.sigma}

The electron self-energy in the $GW$ approximation is:\cite{hl86}
  \begin{equation} \label{eq.sigma}
  \Sigma(\r,\rp;\w) = \frac{i}{2\pi} \! \int_{-\infty}^{+\infty} \!\!\!\!d\wp 
    G(\r,\rp,\w+\wp) W(\r,\rp,\wp) e^{-i\delta\wp},
  \end{equation}
where $\delta$ is a positive infinitesimal. 
At large frequencies the Green's function decays as $\w^{-1}$ and 
the screened Coulomb interaction tends to the frequency-independent
bare Coulomb interaction $v$. As a consequence, the integrand in Eq.\ (\ref{eq.sigma}) 
decays as $\w^{-1}$ and the integration requires some care.

It is convenient to split the self-energy into an exchange contribution 
$\Sigma^{\rm ex}=Gv$ and a Coulomb term $\Sigma^{\rm c}=G(W-v)$.\cite{blochl}
It is easy to check that the integrand in the Coulmb term decays as 
$\w^{-2}$ at large frequencies, therefore the integral is well behaved and the 
integration can be performed by using a numerical cutoff $|\wp|< \wc$ 
in Eq.\ (\ref{eq.sigma}).
A detailed analysis of the analytic properties of the Coulomb term
$\Sigma^{\rm c}(\w)$ shows that it must decay as $\w^{-1}$ 
at large frequencis, and that the use of the cutoff $\wc$ in the integration 
introduces an error of the order of $\w_{\rm p}/\wc$, where $\w_{\rm p}$ denotes
the characteristic plasmon frequency of the system.

In order to integrate the exchange term we observe that 
$\Sigma^{\rm ex}=G^{\rm A}v + G^{\rm N}v$
and that the poles of $G^{\rm A}$ lie in the lower half of the complex plane,
hence the integration of the term $G^{\rm A}v$ yields a vanishing contribution.
On the other hand, the integration of $G^{\rm N}v$ yields a constant (frequency-independent) term.\cite{note.integral}
In summary, we perform the frequency integration in Eq.\ (\ref{eq.sigma}) 
by evaluating numerically
the Coulomb term using an energy cutoff, and by integrating analitycally the exchange term:
  \begin{equation}
  \Sigma(\r,\rp;\w) = \Sigma^{\rm c}(\r,\rp;\w) + \Sigma^{\rm ex}(\r,\rp),
  \end{equation}
with
  \begin{equation}\label{eq.convolution}
 \Sigma^{\rm c}(\r,\rp\!;\w) \! = \!\! \frac{i}{2\pi} \!\!\int_{_{-\wc}}^{^\wc} \!\!\!\!\!\!\!d\wp 
 G(\r,\rp\!;\w+\wp) \!\Big[W(\r,\rp\!,\wp)-v(\r,\rp)\!\Big]\!,
  \end{equation}
and
  \begin{equation}
  \Sigma^{\rm ex}(\r,\rp) = - \sum_v \psi_v^\star(\r)\psi_v(\rp) v(\r,\rp).
  \end{equation}

\section{Implementation in a basis of planewaves}\label{sec.theory.g}

We here describe our planewaves implementation of the scheme 
developed in Sec.\ \ref{sec.theory}. The choice of a planewaves representation
was motivated by the need for making contact with existing literature on dielectric
matrices,\cite{cpm,hl86-prb,balde_tosa,baroni-resta} 
and by the availability of DFPT software for lattice-dynamical
calculations\cite{espresso} which was used as a reference for our implementation.

We adopt the following conventions for the transformation from real to reciprocal space.
The wavefunctions transform as usual according to:
  \begin{equation}
  \psi_{n\k}(\rp)=\frac{1}{\sqrt{\Omega}}\sum_\Gp
  {\rm e}^{i(\k+\Gp)\cdot\rp} u_{n\k}(\Gp),
  \end{equation}
with $\Omega$ the volume of the unit cell and $\k$ the Bloch wavevector.
The bare Coulomb interaction transforms according to
  \begin{equation}\label{eq.v}
  v(\r,\rp) = \frac{1}{N_\q\Omega}  \sum_{\q\G} v(\q+\G)
  {\rm e}^{i(\q+\G)\cdot(\rp-\r)},
  \end{equation}
where $\q$ is also a Bloch wavevector and $N_\q$ is the number of such wavevectors
in our discretized Brillouin zone. In Eq.\ (\ref{eq.v}) we have 
$v(\q+\G) = 4\pi e^2/|\q+\G|^2$.
The latter expression for $v(\q+\G)$ is arrived at by replacing
the integration $\int d\r \exp(i\q\cdot\r)/|\r|$ over the crystal volume by 
an integration over all space. This choice corresponds to assuming that 
we can rely on a very fine sampling of the Brillouin zone.
Had we performed the integration on a sphere
with a radius $R_{\rm c}$ defined by the crystal volume 
($4/3\pi R_{\rm c}^3 = N_\q \Omega$), then we would have obtained
  \begin{equation}
  v_{\rm t} (\q+\G) = \frac{4\pi e^2}{|\q+\G|^2} (1-\cos |\q+\G| R_{\rm c}),
  \end{equation}
which corresponds to the truncated Coulomb potential introduced in Ref.\ \onlinecite{alavi}.
We will come back to this aspect in Sec.\ \ref{sec.coulomb.g}.
The screened Coulomb interaction and the non interacting Green's function
transform according to
  \begin{equation}\label{eq.w.bloch}
  W(\r,\rp;\w) = \!\!\frac{1}{N_\q\Omega} \!\! \sum_{\q\G\Gp} 
  {\rm e}^{-i(\q+\G)\cdot\r}
  W_{\G\Gp}(\q;\w)
  {\rm e}^{i(\q+\Gp)\cdot\rp},
  \end{equation}
and
  \begin{equation}
  G (\r,\rp;\w) = \frac{1}{N_\k\Omega}  \sum_{\k\G\Gp} 
   {\rm e}^{-i(\k+\G)\cdot\r} 
   G_{\G\Gp}(\k;\w)
   {\rm e}^{i(\k+\Gp)\cdot\rp},
  \end{equation}
with similar expressions for $G^A$, $G^N$.
We note that the sign convention adopted here in the Fourier transforms
[e.g.\ ${\rm exp}(+i\q\cdot\rp)$ in the rhs of Eq.\ (\ref{eq.w.bloch})]
is necessary to obtain the compact expression Eq.\ (\ref{eq.deltan.bloch}) below
for the induced charge, and is opposite to the convention adopted in Ref.~\onlinecite{hl86}.
Before proceeding it is also convenient to introduce the ``right-sided'' inverse dielectric matrix
through
  \begin{equation}
  W(\r,\rp;\w) = \int d\rpp v(\r,\rpp) \, \epsilon^{-1}(\rpp,\rp;\w).
  \end{equation}
By adopting the same convention for the inverse dielectric matrix
as for the screened Coulomb interaction 
the above equation can be rewritten as:
  \begin{equation}\label{eq.Wg}
  W_{\G\Gp}(\q;\w) = v(\q+\G)  \epsilon^{-1}_{\G\Gp}(\q;\w).
  \end{equation}
We note that our Eq.\ (\ref{eq.Wg}) is slightly different from the
standard expression [e.g.\ Eq.\ (22) of Ref.\ \onlinecite{hl86}], due
to our choice of using the right-sided inverse dielectric matrix.

\subsection{Screened Coulomb interaction}\label{sec.coulomb.g}

In order to rewrite Eqs.\ (\ref{eq.deltan})-(\ref{eq.w.dfpt}) in
the Bloch representation and in reciprocal space we proceed as follows.
We first write the linear systems Eq.\ (\ref{eq.linsys.1})
by relabeling the wavefunctions $\psi_v$ as Bloch states $\psi_{v\k}$:
  \begin{equation}\label{eq.linsys.3}
  (\H-\E_{v\k}\pm\w) \Delta \psi^\pm_{v\k[\r,\w]}  = -(1-\P)  \Delta V_{[\r,\w]} \psi_{v\k}.
  \end{equation}
The linear variations of the wavefunctions can be expanded in terms
of Bloch waves as follows:
  \begin{equation}\label{eq123}
  \Delta \psi^\sigma_{v\k[\r,\w]} = \frac{1}{N_\q\Omega} \sum_{\q\G} \Delta u^\sigma_{v\k[\q,\G,\w]}
  {\rm e}^{i(\k+\q)\cdot\rp} {\rm e}^{-i(\q+\G)\cdot\r},
  \end{equation}
where $\Delta u^\sigma_{v\k[\q,\G,\w]}$ is cell-periodic in $\rp$.
From the linear variations of the wavefunctions we construct the
induced charge density using Eq.\ (\ref{eq.deltan}):
  \begin{equation}\label{eq.ch}
  \Delta n_{[\r,\w]} = \frac{2}{N_\k}\sum_{v\k\s} \psi_{v\k}^\star  \Delta \psi^\s_{v\k[\r,\w]}.
  \end{equation}
Here the factor $N_\k$ takes into account the normalization of
the Bloch states in the unit cell [the 
wavefunctions $\psi_v$ in Eq.\ (\ref{eq.linsys.1}) are normalized 
in the whole crystal].
Next we expand the screened Coulomb interaction in terms
of Bloch waves: 
  \begin{equation}\label{eq.dv.bloch}
  \Delta V_{[\r,\w]} (\rp) = \frac{1}{N_\q\Omega}  \sum_{\q\G} \Delta v_{[\q,\G,\w]}(\rp) 
   {\rm e}^{-i(\q+\G)\cdot\r} {\rm e}^{i\q\cdot\rp}, 
  \end{equation}
where $\Delta v_{[\q,\G,\w]}$ is cell-periodic in $\rp$.
If we now place Eqs.\ (\ref{eq.dv.bloch}) and (\ref{eq123}) into Eq.\ (\ref{eq.linsys.3})
we discover that the component $\Delta v_{[\q,\G,\w]}$ of the
perturbing potential corresponding to the Bloch wave ${\rm exp}(-i\q\cdot\r)$
couples only to the variations of the wavefunctions corresponding to
the Bloch wave ${\rm exp}[i(\k+\q)\cdot\rp]$.
As a result the linear system Eq.\ (\ref{eq.linsys.1}) becomes:
  \begin{equation}\label{eq.linsys.4}
  (\H_{\k+\q}-\E_{v\k}\pm\w) \Delta u^\pm_{v\k[\q,\G,\w]}  = -(1-\P^{\k+\q}) \Delta v_{[\q,\G,\w]} u_{v\k},
  \end{equation}
where $\H_\k= {\rm e}^{-i\k\cdot\r} \H {\rm e}^{i\k\cdot\r}$ and
$\P^\k=\sum_v | u_{v\k} \rangle \langle u_{v\k}|$.
The induced charge density associated with the Bloch wave ${\rm exp}(-i\q\cdot\r)$ 
now reads 
  \begin{equation} \label{eq.deltan.bloch}
  \Delta n_{[\q,\G,\w]} = \frac{2}{N_\k}\sum_{v\k\s} u_{v\k}^\star  \Delta u^\s_{v\k[\q,\G,\w]}.
  \end{equation}
This result is very similar to the case of standard DFPT.\cite{baroni.rmp} 
The main difference is that in the present case the translational invariance 
of the screened Coulomb interaction induces a coupling between the perturbation 
with Bloch wave ${\rm exp}(-i\q\cdot\r)$ in the variable $\r$ and its induced
response with Bloch wave ${\rm exp}(\,+i\q\cdot\rp)$ in the variable $\rp$.
To conclude our derivation, we rewrite the screened Coulomb interaction 
Eq.\ (\ref{eq.w.dfpt}) after expanding the cell-periodic function 
$\Delta n_{[\q,\G,\w]}(\rp)$ in planewaves:
  \begin{equation}\label{eq.w.deltan.g}
  W_{\G\Gp}(\q;\w) = [\delta_{\G\Gp} + \Delta n_{[\q,\G,\w]}(\Gp) ]v(\q+\Gp).
  \end{equation}

In practical calculations we proceed as follows: we first initialize the perturbation
in the linear systems using 
$\Delta v^{\rm bare}_{[\q,\G,\w]}(\rp) = v(\q+\G) \, {\rm exp}(\,i\G\cdot\rp)$.
The solution of the linear systems yields the change in the wavefunctions,
which are then used to construct the induced charge, the induced Hartree potential,
and the updated screened Coulomb interaction. We repeat this procedure by starting with
the updated screened Coulomb interaction until convergence is achieved.
At convergence the self-consistent perturbing potential yields
the screened Coulomb interaction $W_{\G\Gp}(\q;\w)$.
The calculation must be repeated for every perturbation, i.e.\ for
each set of parameters $[\q,\G,\w]$.
At the end of this procedure it is straightforward to obtain the inverse dielectric matrix 
$\epsilon^{-1}_{\G\Gp}(\q;\w)$ using Eqs.~(\ref{eq.w.deltan.g}) and 
(\ref{eq.Wg}). Alternatively, it is also possible to scale the initial
perturbation and use $\Delta v^{\rm bare}_{[\q,\G,\w]} = {\rm exp}(i\G\cdot\rp)$
to obtain the inverse dielectric matrix at the end of the self-consistent procedure
[indeed Eq.~(\ref{eq.linsys.4}) is a linear system].

The scheme developed here allows us to calculate one row (in $\Gp$) of the inverse 
dielectric matrix $\epsilon^{-1}_{\G\Gp}(\q;\w)$ by determining the linear response to the
perturbation ${\rm exp}[-i(\q+\G)\cdot\r]$. This idea has been discussed already
in Ref.\ \onlinecite{kunc-tosatti} in the framework of nonperturbative methods
based on supercell calculations.

\subsubsection{Singularities in the inverse dielectric matrix and the screened Coulomb interaction}

In order to avoid the singular behavior of the wings of the inverse dielectric matrix 
in the long wavelength limit ($|\q|\!\rightarrow\!0$) it is convenient to work with
the symmetrized inverse dielectric matrix defined as follows:\cite{balde_tosa}
  \begin{equation}
  \tilde\epsilon^{-1}_{\G\Gp}(\q;\w) = \epsilon^{-1}_{\G\Gp}(\q;\w)  \frac{|\q+\Gp|}{|\q+\G\phantom{^\prime}|}.
  \end{equation}
Unlike its unsymmetrized counterpart $\epsilon^{-1}_{\G\Gp}(\q;\w)$,
the wings of $\tilde\epsilon^{-1}_{\G\Gp}(\q;\w)$ have finite limits 
at long wavelengths.
The screened Coulomb interaction of Eq.\ (\ref{eq.Wg}) is now rewritten in symmetrized form
as
  \begin{equation}\label{eq.Wg_symm}
  W_{\G\Gp}(\q;\w) = \frac{4\pi e^2}{|\q+\G||\q+\Gp|}  \tilde\epsilon^{-1}_{\G\Gp}(\q;\w).
  \end{equation}
While the symmetrized inverse dielectric matrix has finite limits at long wavelengths,
the screened Coulomb interaction still presents a divergence corresponding to the
long-range tail of the Coulomb potential in real space. This divergence requires special
handling when performing the Brillouin zone integration to calculate 
the $GW$ self-energy.\cite{hl86}
We here overcome this difficulty following the prescription of
Ref.~\onlinecite{alavi}. For this purpose we replace
the bare Coulomb potential $v(\r,\rp)$ by the truncated potential 
$v_{\rm t}(\r,\rp)=v(\r,\rp)[1-\Theta(|\r-\rp|-R_{\rm c})]$, 
$\Theta(x)$ being the Heaviside step function.
The truncation radius is defined as in Sec.\ \ref{sec.coulomb.g}. 
Using this truncated Coulomb potential, the final expression for the screened 
Coulomb interaction in reciprocal space becomes:
  \begin{equation}
  W^{\rm t}_{\G\Gp}(\q;\w) = 4\pi e^2 \, \frac{1-\cos{R_{\rm c}|\q+\G|}}{|\q+\G||\q+\Gp|}\,
     \tilde\epsilon^{-1}_{\G\Gp}(\q;\w).
  \end{equation}
In the long-wavelength limit $\q\!\rightarrow\!0$ the head 
of the truncated screened Coulomb interaction ($\G=\Gp=0$) tends to
the finite limit 
$2\pi e^2 \, R_{\rm c}^2 \, \tilde\epsilon^{-1}_{00}(\q\!\rightarrow\!0;\w)$
and the singular behavior is removed.
Optimized truncation strategies have been developed for non-isotropic materials
and systems with reduced dimensionality.\cite{sohrab}

\subsection{Green's function}\label{sec.green.g}

We now specialize Eqs.\ (\ref{eq.green.1})-(\ref{eq.green.3}) to the case
of a planewaves basis and the Bloch representation.
We start by rewriting Eq.\ (\ref{eq.green.1}) after relabeling the electronic states
$\psi_n$ as Bloch states $\psi_{n\k}$ and taking into account the
normalization, as already done in Sec.\ \ref{sec.coulomb.g}.
Next we expand the Green's function in terms
of the Bloch waves ${\rm exp}[-i(\k+\G)\cdot\r]$ and ${\rm exp}(i\k\cdot\rp)$:
  \begin{equation}\label{eq.green.bloch}
  G^{\rm A}_{[\r,\w]} (\rp) = \frac{1}{N_\k\Omega}  \sum_{\k\G} g^{\rm A}_{[\k,\G,\w]}(\rp)
   {\rm e}^{-i(\k+\G)\cdot\r} {\rm e}^{i\k\cdot\rp},
  \end{equation}
with $g^{\rm A}_{[\k,\G,\w]}(\rp)$ cell-periodic in $\rp$. An analogous expansion
holds for the non-analytic component $G^{\rm N}$.
Equations (\ref{eq.green.split.2}),(\ref{eq.green.split.3}) are now rewritten as:
  \begin{equation}\label{eq.green.3.g}
   (\H_\k-\w^+)  g^{\rm A}_{[\k,\G,\w]}(\Gp)  =  -\,\delta_{\G\Gp},
  \end{equation}
  \begin{equation} \label{eq.green.4b.g}
  g^{\rm N}_{[\k,\G,\w]}(\Gp)  =  
  2\pi i \sum_v \delta(\w-\E_{v\k}) u^\star_{v\k}(\G)u_{v\k} (\Gp).
  \end{equation}
In deriving Eqs.\ (\ref{eq.green.3.g}),(\ref{eq.green.4b.g}) we made use once again
of time-reversal symmetry, yielding
$u^\star_{v\k}(\G) = u_{v,-\k}(-\G)$.
Similarly to the case of the screened Coulomb interaction, by solving the
linear system in Eq.\ (\ref{eq.green.3.g}) for a set of parameters $[\k,\G,\w]$ 
we obtain an entire row $\Gp$ of the analytic component of the
Green's function $g^{\rm A}_{[\k,\G,\w]}(\Gp)$.

\subsection{Self-energy}\label{sec.sigma.g}

The $GW$ self-energy in Eq.\ (\ref{eq.sigma}) is calculated in real space
after performing the Fourier transforms of $W_{\G\Gp}(\q;\w)$ and $G_{\G\Gp}(\k;\w)$.
The result is then transformed back in reciprocal space to obtain
$\Sigma_{\G\Gp}(\k;\w)$. The evaluation of the matrix elements of the
self-energy in the basis of Kohn-Sham eigenstates is performed in reciprocal space.
Since the planewaves cutoff required to describe the inverse dielectric
matrix and the self-energy is typically much smaller than the cutoff used
in density-functional calculations,\cite{hl86} the procedure described here 
does not require an excessive computational effort and accounts for only
a fraction of the total computation time.

\section{Scaling properties}\label{sec.scaling}

In this section we analyze the computational complexity of the algorithms
proposed in Sec.\ \ref{sec.theory.g}, by focusing on our planewaves
implementation. Without loss of generality we consider a $\Gamma$ point sampling 
of the Brillouin zone and we leave aside the frequency-dependence.
We assume that the Kohn-Sham electronic wavefunctions are expanded in a basis of planewaves
with a kinetic energy cutoff $E_{\rm cut}^{\rm wf}$, corresponding to
$N_\G^{\rm wf}$ plane waves.
In the simplest case of norm-conserving pseudopotential approaches 
the electronic charge density is described using a basis set with a cutoff
$E_{\rm cut}^{\rm den}=4E_{\rm cut}^{\rm wf}$, and the corresponding numbers 
of basis functions and real-space grid points are $N_\G^{\rm den}$ and
$N_\r^{\rm den}$, respectively. The screened Coulomb interaction and the Green's function
are described by a smaller cutoff $E_{\rm cut}^{\rm s}$ and $N_\G^{\rm s}$
planewaves. The self-energy is expanded in a planewaves basis with cuttoff
$E_{\rm cut}^{\rm SE}=4E_{\rm cut}^{\rm s}$, and we denote by $N_\r^{\rm SE}$
the number of real-space grid points associated with this basis.

\subsubsection{Screened Coulomb interaction}\label{sec.coulomb.scaling}

Equation (\ref{eq.linsys.4}) needs to be solved for each one of the $N_\G^{\rm s}$ planewave perturbations
and the $N_v$ occupied electronic states. For the solution
of Eq.~(\ref{eq.linsys.4}) we adopt the complex bi-orthogonal conjugate gradient method
of Ref.~\onlinecite{jacobs} (cBiCG), as described in Appendix \ref{app.cbcg}.
Each solution of Eq.\ (\ref{eq.linsys.4}) requires two cBiCG minimizations
(for $\pm\w$), and each cBiCG minimization consists of two conjugate gradients (CG) sequences.
The most time-consuming operation in each CG step is the application of the Hamiltonian 
to the previous search direction, and in particular
the Fourier transform of the wavefunctions to real-space and back for evaluating
the product with the local potential. Fast-Fourier-transform (FFT)
algorithms allow us to perform these calculation in $N^{\rm den}_{\rm FFT} = 4 N_\r^{\rm den} {\rm log} N_\r^{\rm den}$ floating point
operations.\cite{frigo}
If in average the CG minimization requires $N_{\rm CG}$ steps
and the self-consistency loop requires $N_{\rm SCF}$ iterations,
then the total cost of the entire calculation corresponds to a number of floating point operations
   \begin{equation}\label{eq.scaling.1}
   N^{{\rm S}GW}_{\rm flops} = 8 N_{\rm CG} N_{\rm SCF} N_\G^{\rm s}  N_v N^{\rm den}_{\rm FFT},
   \end{equation}
where S$GW$ stands for ``Sternheimer GW''.
As $N_\G^{\rm s}$, $N_v$, and $N_\r^{\rm den}$ scale linearly with the size of the system
as measured by the number of atoms $N_{\rm at}$,
the overall scaling of this procedure is $N_{\rm at}^3 {\rm log} N_{\rm at}$.

For comparison it is useful to consider the scaling of standard $GW$ calculations
based on the expansion over unoccupied states (hereafter referred to as the ``HL'' method).\cite{hl86} 
The calculation of the irreducible RPA polarization requires the evaluation of the 
optical matrix elements between each of the $N_v$ occupied states and each of the $N_c$ unoccupied states.
These matrix elements are typically computed by using Fourier transforms of $\psi_c^\star(\r)\psi_v(\r)$,
therefore this procedure requires essentially $N_v N_c$ Fourier transforms from real- to reciprocal-space.
Each Fourier transform is performed on the real-space grid for the density with $N_\r^{\rm den}$ grid points, 
therefore the total cost of the standard method corresponds to a number of floating point operations
   \begin{equation}\label{eq.scaling.2}
   N^{\rm HL}_{\rm flops} = N_c  N_v N^{\rm den}_{\rm FFT}.
   \end{equation}
Even in this case therefore the overall scaling is $N_{\rm at}^3 {\rm log} N_{\rm at}$.

Since the method of Ref.\ \onlinecite{hl86} calculates the dielectric matrix and then performs a matrix inversion,
in order to compare the prefactors in Eqs.\ (\ref{eq.scaling.1}) and (\ref{eq.scaling.2})
we consider the non self-consistent calculation of the dielectric matrix as described in Sec.\ \ref{sec.diel.nscf}
[$N_{\rm SCF}=1$ in Eq.\ (\ref{eq.scaling.1})], and we restrict ourselves to 
the calculation of the static dielectric matrix.
In this case only one calculation of Eq.\ (\ref{eq.linsys.1}) is required instead of two for $\pm\w$, 
and the two CG sequences of the cBiCG algorithm do coincide.
As a result, a factor 4 drops out of the prefactor in Eq.\ (\ref{eq.scaling.1}).
If we assume for definiteness a perfectly well-conditioned linear system (condition number $\kappa=1$),
and express the number of CG iterations required to achieve convergence through Eq.\ (\ref{eq.cg}),
then the ratio between the complexity of the S$GW$ approach in a planewaves
implementation and the standard approach becomes
   \begin{equation}\label{eq.ratio.flops}
   N^{{\rm S}GW}_{\rm flops}/N^{\rm HL}_{\rm flops} = N_\G^{\rm s}/N_c {\rm log}(2/\varepsilon),
   \end{equation}
where $\varepsilon$ is the relative accuracy of the results.
As an example, for a relative accuracy of $\varepsilon=10^{-5}$ 
we find this ratio to be $\simeq 12 N_\G^{\rm s}/N_c$.
In the case of silicon, using a typical cutoff $E_{\rm cut}^{\rm s}=10$ Ry we obtain $N_\G^{\rm s}=137$, therefore
the S$GW$ approach becomes convenient when more than $\sim$1650 unoccupied states are used
in the standard approach.
This is rarely the case as most calculations reported to date use only a few hundreds of 
unoccupied electronic states.
Of course the accuracy of the standard sum-over-states expression is difficult to
quantify, and probably a convergence on 5 significant digits is not warranted by a few
hundreds of electronic states. 

Our estimate suggests that the planewaves implementation of our method
can be as expensive as the standard approach.
It should be noted, however, that our scheme has the advantage of providing
the whole self-energy $\Sigma(\r,\rp;\w)$, while the standard approach 
typically provides the matrix elements of the self-energy on a small subset of states of the order of $N_v$. 
Therefore if we were to perform a comparison based on the
same amount of output information, we should use $N_c N_\G^{\rm s}$ in Eq.\ (\ref{eq.scaling.2})
instead of $N_c N_v$. In this case Eq.\ (\ref{eq.ratio.flops}) would change into
  \begin{equation}
  N^{{\rm S}GW}_{\rm flops}/N^{\rm HL}_{\rm flops} = N_v /N_c{\rm log}(2/\varepsilon),
  \end{equation}
and for $\varepsilon=10^{-5}$ we would have $N^{{\rm S}GW}_{\rm flops}/N^{\rm HL}_{\rm flops}\simeq 12 N_v/N_c$.
This clearly shows that, if the entire self-energy was needed (as opposed to a few matrix elements),
then our proposed S$GW$ approach would be more convenient that the standard sum-over-states
approach.

The above analysis shows that the main bottlenecks of our method 
are (i) the Fourier transform for the application of the Hamiltonian and (ii) the large basis sets adopted.
In order to make the approach proposed here more efficient we could either
move to real-space methods where the application of the Hamiltonian scales linearly with system size,\cite{chelikowsky}
or reduce the size of the basis set by using local orbitals.\cite{siesta} 
Fast evaluations of the operation $\hat{H}\psi$ in order-$N$ operations should 
make it feasible $GW$ calculations with $N_{\rm at}^3$ scaling and with a very favorable
prefactor. We will come back to this aspect in Sec.\ \ref{sec.conclusions}.

\subsubsection{Green's function}\label{sec.green.scaling}

The complexity of the procedure for calculating the Green's function proposed
in Sec.\ \ref{sec.green} can be analyzed along the same
lines of Sec.\ \ref{sec.coulomb.scaling}. The main differences in this case are that (i) the linear system
Eq.\ (\ref{eq.green.3}) does not depend on the occupied states,
(ii) the calculation is non self-consistent, and (iii) the calculation is performed
for one single frequency $\w^+$, while the entire frequency-dependence is generated
through the multishift method. As a result, a factor $2 N_{\rm SCF} N_v$ drops out
of Eq.\ (\ref{eq.scaling.1}) and the computational cost of the Green's function
calculation reads:
   \begin{equation}\label{eq.scaling.3}
   N^{\rm GF}_{\rm flops} = 4 N_{\rm CG} N_\G^{\rm s} N^{\rm den}_{\rm FFT}.
   \end{equation}
The complexity of this calculation is significantly smaller than the complexity
of the algorithm for the screened Coulomb interaction. In particular, the
calculation of the Green's function scales as $N_{\rm at}^2 {\rm log} N_{\rm at}$.
This procedure for calculating the Green's function is advantageous
with respect to an expansion over empty states, as the orthogonalization
of the unoccupied states would require a number of floating point operations 
scaling as $\sim N_{\rm at}^3$.

\subsubsection{Scaling of the self-energy calculation}\label{sec.sigma.scaling}

The self-energy is computed in real-space after obtaining $G(\r,\rp;\w)$ and $W(\r,\rp;\w)$ 
from $G(\G,\Gp;\w)$ and $W(\G,\Gp;\w)$, respectively, and then is tranformed back into reciprocal space.
The 6-dimensional FFT transforms require $(N_\G^{\rm s}+N_\r^{\rm SE})N^{\rm SE}_{\rm FFT}$
operations for each frequency of the screened Coulomb interaction,
having defined $N^{\rm SE}_{\rm FFT}=4 N_\r^{\rm SE} {\rm log} N_\r^{\rm SE}$.
The computational cost of this procedure scales as $N_{\rm at}^2 {\rm log} N_{\rm at}$,
and is small with respect to the cost of calculating the screened
Coulomb interaction.

\section{Results}\label{sec.results}

In order to demonstrate the approach proposed in Secs.~\ref{sec.theory},\ref{sec.theory.g}  we have realized a prototype implementation
within the empirical pseudopotential method (EPM) of Ref.\ \onlinecite{cohen_berg},
and we have validated our implementation for the test case of silicon.

\subsection{Dielectric matrix}\label{sec.5a}

Table \ref{tab.1} contains some of the components of the inverse dielectric matrix calculated using
the self-consistent Sternheimer method described in Sec.~\ref{sec.theory.g}.
In all our calculations we used inverse dielectric matrices of size 59$\times$59, 
corresponding to a kinetic energy cutoff of 5 Ry for the screened Coulomb interaction.
For the purpose of comparison with Ref.\ \onlinecite{balde_tosa} 
we calculated the static and long wavelength limit ($\w=0$, $\q \rightarrow 0$) of the inverse dielectric matrix
for the first few reciprocal lattice vectors. 
The authors of Ref.\ \onlinecite{balde_tosa} adopted the standard approach based 
on the expansion over the unoccupied electronic states of the dielectric matrix, 
and obtained the inverse dielectric matrix by performing matrix inversions.
In our calculations we used the self-consistent method of Sec.~\ref{sec.theory.g}
and no matrix inversion was necessary.
The excellent agreement which can be seen in Table \ref{tab.1} between 
our calculations and Ref.~\onlinecite{balde_tosa} supports the validity of our approach.

\begin{table}
\caption{\label{tab.1} Long-wavelength limit of the static symmetrized
inverse dielectric matrix of silicon $\tilde\epsilon^{-1}_{\G\Gp}(\q;\w)$
[$\q=(0.01,0,0)2\pi/a$ and $\w=0$]. We compare our calculations performed within
the self-consistent Sternheimer approach
with the results obtained in Ref.\ \onlinecite{balde_tosa}
using the expansion over empty states and the inversion of the dielectric matrix.
For the calculations we sampled the Brillouin zone with 29 irreducible $\k$-points,
corresponding to a $8\times8\times8$ grid,\cite{balde_tosa,baroni-resta}
and a plane wave cutoff of 5 Ry.\cite{balde_tosa} Following Ref.\ \onlinecite{balde_tosa}
we employed the empirical pseudopotential parameters from Ref.\ \onlinecite{cohen_berg}.
The reciprocal lattice vectors are in units of $2\pi/a$, $a$ being the
lattice parameter.
\vspace{0.5cm}}
\begin{tabular}{c c c}
\hline
\hline
   & \multicolumn{2}{c}{$\tilde\epsilon^{-1}_{\G\Gp}(\q;\w)$}  \\
   \multicolumn{3}{c}{\vspace{-0.4cm}}  \\
$\,\,\G$\phantom{ciao} $\,\Gp$   & Ref.\ \onlinecite{balde_tosa}  &  Present work \\
\hline
    (0,0,0) (0,0,0)   & \phantom{-}0.083    &  \phantom{-}0.0866  \\
 (1,1,1)  (1,1,1)     &   \phantom{-}0.605  & \phantom{-}0.6055 \\
(\mo,1,1) (1,1,1)     &   \phantom{-}0.008  & \phantom{-}0.0076 \\
 (1,\mo,1) (\mo,1,1)  & \phantom{-}0.010    & \phantom{-}0.0102 \\
 (1,\mo,\mo) (\mo,1,1)& \phantom{-}0.045    & \phantom{-}0.0463 \\
 (2,0,0) (1,1,1)      &    -0.038           & -0.0382 \\
 (2,0,0) (\mo,1,1)    &    -0.005           & -0.0049 \\
 (2,0,0) (2,0,0)      &  \phantom{-}0.667   & \phantom{-}0.6671 \\
 (\mt,0,0) (2,0,0)    & \phantom{-}0.006    & \phantom{-}0.0063 \\
 (0,2,0) (2,0,0)      &  \phantom{-}0.016   & \phantom{-}0.0166 \\
\hline
\hline
\end{tabular}
\end{table}

\begin  {figure}
\begin  {center}
\includegraphics[width=7.5cm]{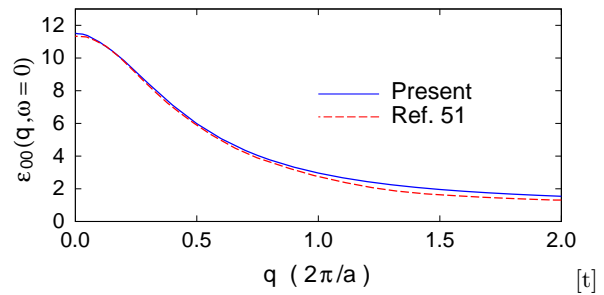}[t]
\end    {center}
\caption{\label{fig1}
        (Color online)
        Dielectric function of silicon calculated using the empirical pseudopotential method and the
        non self-consistent Sternheimer method of Sec.\ \ref{sec.diel.nscf}:
        calculated head of the static dielectric function as a function of wavevector $\epsilon_{00}(\q,\w=0)$ (blue solid line),
        and results from Ref.\ \onlinecite{cohen.diel1} (red dashed line). We used a planewave kinetic energy cutoff of 5 Ry
        and sampled the Brillouin zone through a uniform $8\times 8\times 8$ grid.
        The wavevectors are in units of $2\pi/a$, $a$ being the lattice parameter.
        }
\end    {figure}

Next we consider the wavevector dependence of the head of the
dielectric matrix $\epsilon_{00}(\q,\w=0)$. We performed the
calculation by using the non self-consistent method described in
Sec.\ \ref{sec.diel.nscf} in order to compare our results with Ref.\ \onlinecite{cohen.diel1}.
Figure \ref{fig1} shows that our calculations are in very good agreement with the results
of Ref.\ \onlinecite{cohen.diel1}. The slight differences between our results and those
of Ref.~\onlinecite{cohen.diel1} at large wavevectors can likely be ascribed to the use of a limited 
number of empty states in the perturbative expansion over unoccupied states in the latter work.

\begin  {figure}
\begin  {center}
\includegraphics[width=6.5cm]{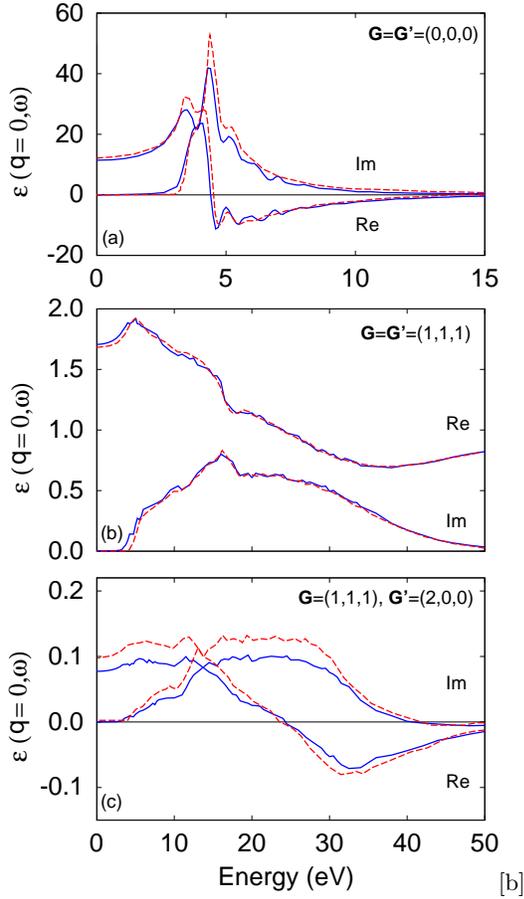}[b]
\end    {center}
\caption{\label{fig2}
        (Color online)
        Frequency-dependent dielectric matrix of silicon at long wavelength [$\q=(0.01,0,0) \, 2\pi/a$].
        The calculations were performed using the empirical pseudopotential method of Ref.\ \onlinecite{cohen_berg}
        and the non self-consistent Sternheimer method of Sec.\ \ref{sec.diel.nscf}.
        We used a planewave kinetic energy cutoff of 5 Ry and sampled the Brillouin zone thorough a uniform
        $24\times 24\times 24$ grid. Such a dense Brillouin zone sampling was necessary to correctly describe
        the absorption onset. In order to avoid null eigenvalues in the linear system Eq.\ (\ref{eq.linsys.1.nscf})
        we performed the calculations by including a small imaginary component of 0.1~eV in the frequency $\w$.
        The panels (a)-(c) correspond to the cases illustrated in Fig.\ 3 of Ref.\ \onlinecite{hl86}
        and show $\epsilon_{\G\Gp}(\q\rightarrow 0,\w)$ for (a) $\G=\Gp=0$, (b) $\G=\Gp=(1,1,1) 2\pi/a$,
        and (c) $\G=(1,1,1) 2\pi/a$, $\Gp=(2,0,0) 2\pi/a$.
        The blue solid lines are our calculations, the red dashed lines are from Ref.\ \onlinecite{hl86}.
        The real and imaginary parts of the dielectric matrix are indicated in each panel.
        We note that the scales on the vertical axes correspond to three different orders of magnitude.
        }
\end    {figure}

Figure \ref{fig2} compares our results for the frequency dependence of the dielectric matrix
at long wavelength with the results reported in Ref.\ \onlinecite{hl86}. 
We focused in particular on the cases illustrated in Fig.\ 3 of Ref.\ \onlinecite{hl86}.
Apart from some small differences possibly arising from the use of the expansion over unoccupied
states in Ref.\ \onlinecite{hl86}, even in this case the agreement between our calculations 
and those of Ref.\ \onlinecite{hl86} is very good throughout the entire frequency range.
The agreement is consistently good for the head of the dielectric matrix and for
diagonal and off-diagonal matrix elements.

\subsection{Pad\'e approximants and convergence}\label{sec.conv}

\begin  {figure}
\begin  {center}
\includegraphics[width=7.5cm]{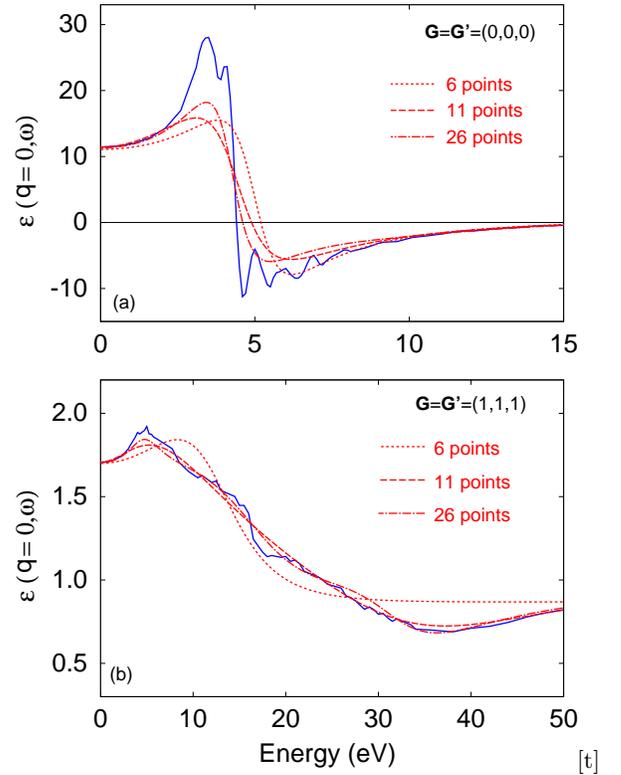}[t]
\end    {center}
\caption{\label{fig3}
        (Color online)
        Real part of the long-wavelength dielectric matrix of silicon as a function of frequency.
        Panels (a) and (b) of this figure correspond to panels (a) and (b) of Fig.~\ref{fig2}, respectively. The
        technical details of the calculations are the same as those described in the caption of Fig.\ \ref{fig2}.
        Solid blue lines: dielectric matrices calculated directly along the real frequency axis,
        from  Fig.\ \ref{fig2}. Dotted, dashed, and dash-dotted red lines: dielectric matrices
        obtained from the analytic continuation on the real axis using Pad\'e approximants of order
        6, 11, and 26, respectively. The Pad\'e approximants were generated using dielectric
        matrices calculated along the imaginary frequency axis on uniform frequency grids
        in the range 0-50 eV. For instance, the 6-points approximant corresponds to calculations
        at the imaginary frequencies of $0,10,\cdots,50$ eV.
        }
\end    {figure}

Figure \ref{fig3} shows the quality of the analytic continuation from the imaginary to
the real frequency axis using Pad\'e approximants (cf.\ Appendix \ref{app.pade}).\cite{pade1,pade2} We found that this procedure based on 
Pad\'e approximants is generally very stable and requires minimal manual intervention.
Approximants of order 5 and higher are able to reproduce the location, the strength, and the width of
the main plasmon-pole structure of the dielectric matrix. Head, wings, and body of the
dielectric matrix are all described consistently (cf.\ Fig.\ \ref{fig3}). Although the singularity
corresponding to the absorption onset in Fig.\ \ref{fig3}(a) is smoothed out by 
Pad\'e approximants of low order, this effect
is washed out when calculating the frequency convolution of the Green's function with
the screened Coulomb interaction for the $GW$ self-energy.
The advantages of performing calculations along the imaginary axis are that (i) the 
linear system in Eq.\ (\ref{eq.linsys.1}) becomes increasingly more well-conditioned
when approaching large imaginary frequencies, and (ii) a moderate Brillouin-zone sampling is
required to perform calculations along the imaginary axis unlike the case of real-axis
calculations. As a result, the worst case scenario for the solution
of the linear system Eq.\ (\ref{eq.linsys.1}) corresponds to the static case $\w=0$.
These technical aspects are described in detail in Appendix~\ref{app.condition}.

The typical number of non self-consistent iterations required to solve Eq.\ (\ref{eq.linsys.1})
for a fixed $\Delta V_{[\r,\w]}$ with a relative accuracy of $\varepsilon_{\rm NSCF}=10^{-10}$
using the cBiCG algorithm described in Appendix \ref{app.cbcg} is $N_{\rm CG}\simeq 21$
(using the preconditioner of Ref.\ \onlinecite{tpa}). This estimate has been obtained by averaging
over all the $\G,\Gp$ reciprocal lattice vectors, $\q$-vectors, and imaginary frequencies.
The typical number of self-consistent cycles required to obtain the screened Coulomb
interaction through Eq.\ (\ref{eq.linsys.1}) with a relative accuracy of $\varepsilon_{\rm SCF}=10^{-5}$ 
is $N_{\rm SCF}\simeq 5$.
Charge-sloshing effects are attenuated by using the potential mixing method proposed in
 Ref.\ \onlinecite{johnson}, appropriately modified to deal with complex potentials. 

For completeness we report here the corresponding figures for
the calculation of the Green's function using Eq.~(\ref{eq.green.3}). The average number
of non self-consistent iterations required to obtain the analytic part of the Green's
function is $N_{\rm GF}\simeq 25$ when preconditioning is adopted (for this purpose 
we used a straightforward adaptation 
of the method of Ref.\ \onlinecite{tpa}). However, multishift minimizations as
described in Appendix \ref{app.multishift} do not allow for the use of preconditioning, and in the
latter case the number of iterations required to achieve convergence (within a relative
accuracy $\varepsilon_{\rm NSCF}=10^{-10}$) can be as high as $N_{\rm GF}\simeq 100$. 

\subsection{Self-energy}\label{sec.5c}

\begin  {figure}
\begin  {center}
\includegraphics[width=6cm]{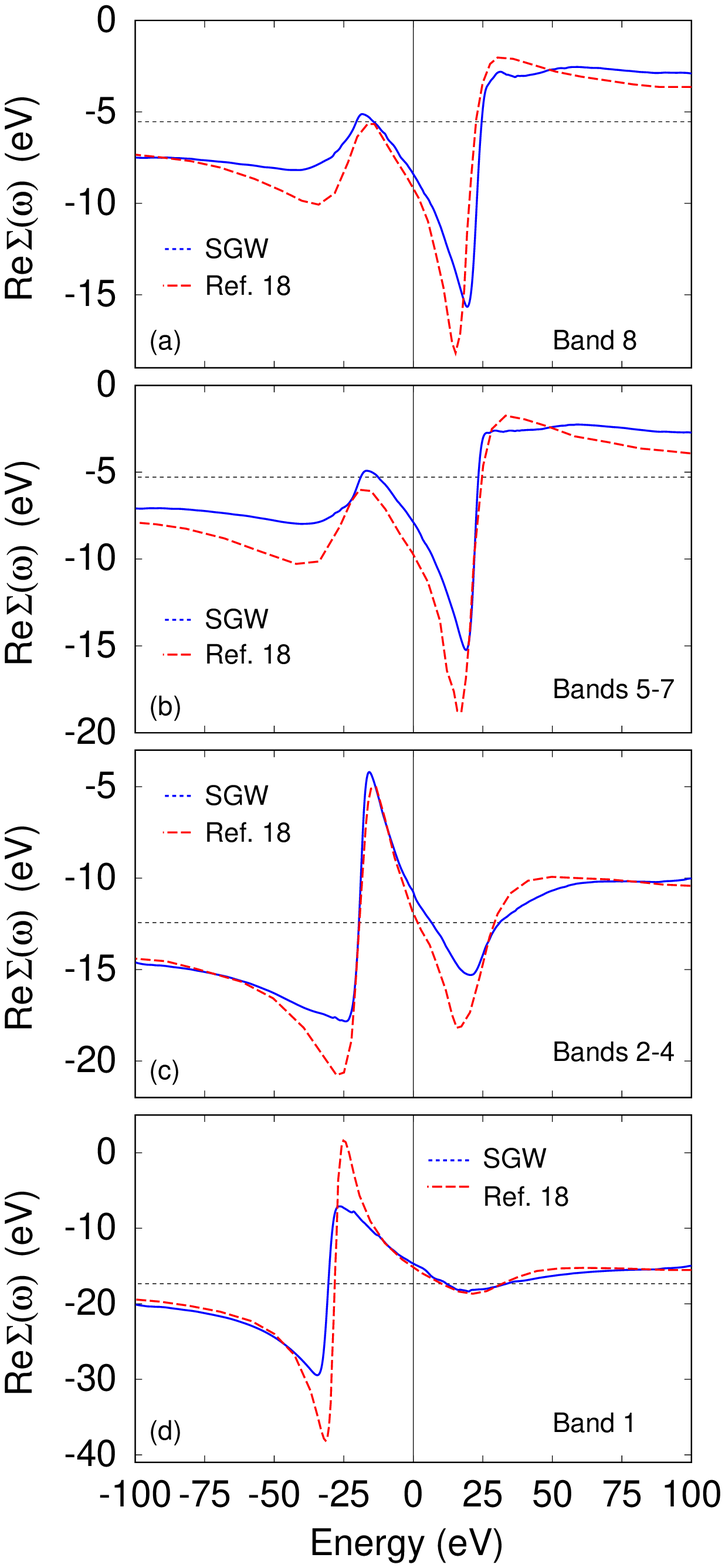}[t]
\end    {center}
\caption{\label{fig4}
        (Color online)
        Real part ${\rm Re} \langle n\k| \Sigma(\r,\rp;\w)|n\k\rangle$   of the expectation value of the $GW$ self-energy for the first
        8 silicon eigenstates at $\k=0$. The solid blue lines are our S$GW$ results using the empirical
        pseudopotential method. The dashed red lines are from the first-principles calculation of Ref.\ \onlinecite{blochl}
        using the LDA and the PAW method. Panels (a)-(d) correspond to the states
        $\Gamma^\prime_{2c}$ (Band 1), $\Gamma_{15c}$ (Bands 2-4), $\Gamma^\prime_{25v}$ (Bands 5-7),
        and $\Gamma_{1v}$ (Band 8), respectively. The energy axis is aligned with the top of the valence band.
        The horizontal dotted lines indicate the calculated bare exchange contribution to the self-energy ${\rm Re} \langle n\k| \Sigma^{\rm ex}(\r,\rp;\w)|n\k\rangle$.
        }
\end    {figure}

\begin  {figure}
\begin  {center}
\includegraphics[width=6cm]{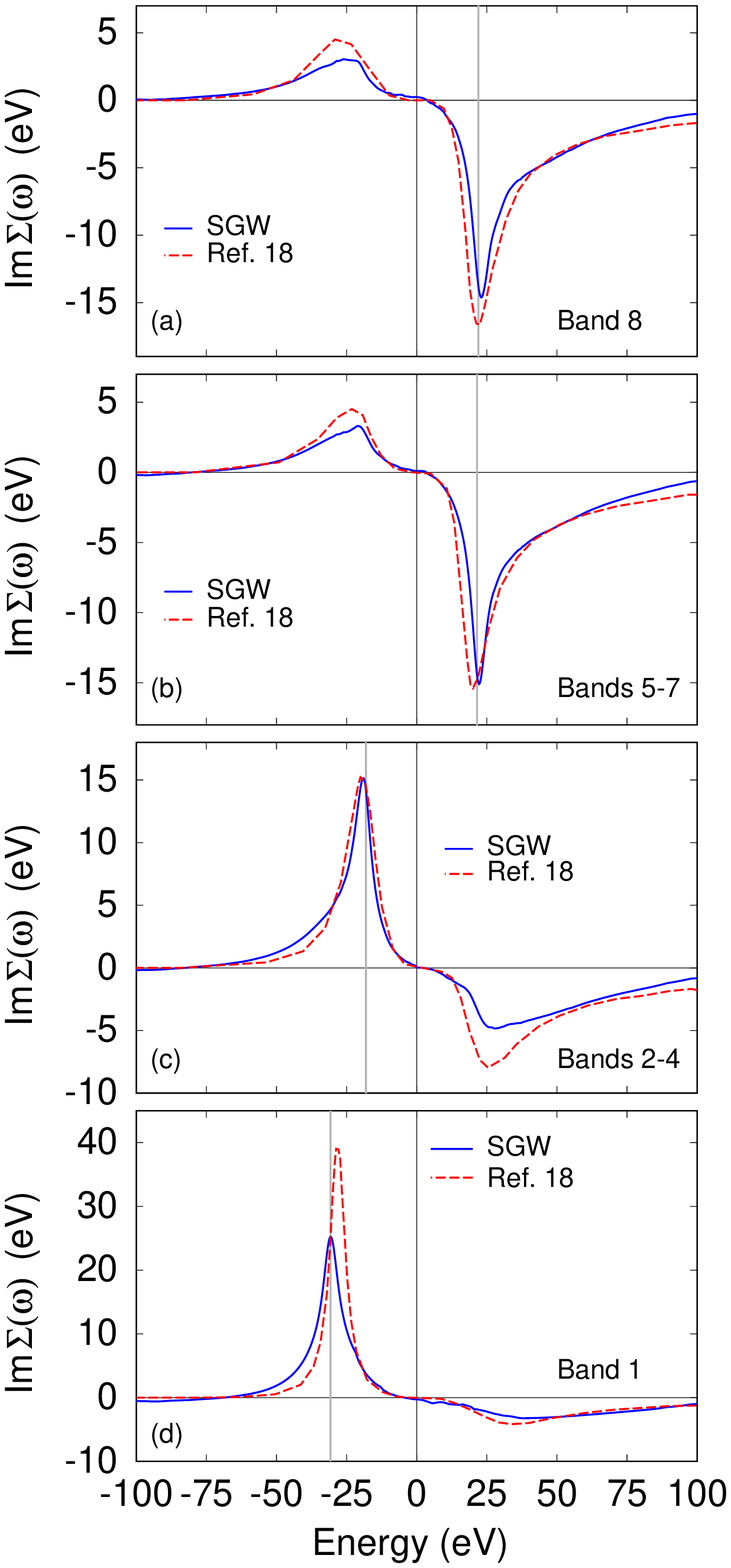}[t]
\end    {center}
\caption{\label{fig5}
        (Color online)
        Imaginary part ${\rm Im} \langle n\k| \Sigma(\r,\rp;\w)|n\k\rangle$ of the expectation value of the $GW$ self-energy for the first
        8 silicon eigenstates at $\k=0$. The solid blue lines are our S$GW$ results using the empirical
        pseudopotential method. The dashed red lines are from the first-principles calculation of Ref.\ \onlinecite{blochl}.
        Panels (a)-(d) correspond to the states $\Gamma^\prime_{2c}$, $\Gamma_{15c}$, $\Gamma^\prime_{25v}$, and $\Gamma_{1v}$, respectively.
        The energy axis is aligned with the top of the valence band.
        For comparison, the vertical grey lines indicate the locations of the logarithmic singularities at $\E_{n\k}\pm\w_{\rm p}$
        ($\w_{\rm p}$ being the plasmon energy) that would arise using a model plasmon-pole dielectric function.\cite{hl}
        }
\end    {figure}

Figures \ref{fig4} and \ref{fig5} show the real part ${\rm Re} \langle n\k| \Sigma|n\k\rangle$ 
and the imaginary part ${\rm Im} \langle n\k| \Sigma|n\k\rangle$ of the
$GW$ self-energy calculated for the first few silicon eigenstates at $\Gamma$
using our S$GW$ method within the EPM scheme. Our results are compared
with the calculations of of Ref.\ \onlinecite{blochl} performed within DFT/LDA 
and the projector-augmented wave method (PAW).
We calculated the screened Coulomb interaction by using a uniform $6\times 6 \times 6$ grid 
to sample the Brillouin zone, and Pad\'e approximants of order 7 along the imaginary
frequency axis. The frequency integration of the Coulomb term $\Sigma^{\rm c}$ of the $GW$ 
self-energy in Eq.~(\ref{eq.convolution}) was performed by using a Coulomb cutoff 
$\wc = 100$ eV and a grid spacing of 0.5 eV. The Green's function was calculated
using an imaginary component $\eta = 0.3$ eV in Eq.\ (\ref{eq.green.3}).
Apart from some differences in the damping of the plasmaron peaks,
the agreement between our calculated self-energy and the results of Ref.~\onlinecite{blochl}
is rather good throughout the entire frequency range $\pm$100 eV.
This finding is quite surprising since we are comparing our empirical pseudopotential
calculations with low kinetic energy cutoff (5 Ry) with the {\it ab-initio} LDA
calculations including PAW core-reconstruction of Ref.\ \onlinecite{blochl}.
Such agreement probably reflects the ability of the EPM method to provide not only
a good description of the band structure of silicon, but also a reasonable
description of the electronic wavefunctions.

\subsection{Quasi-particle excitations and spectral function}\label{sec.5d}

\begin  {figure}
\begin  {center}
\includegraphics[width=6cm]{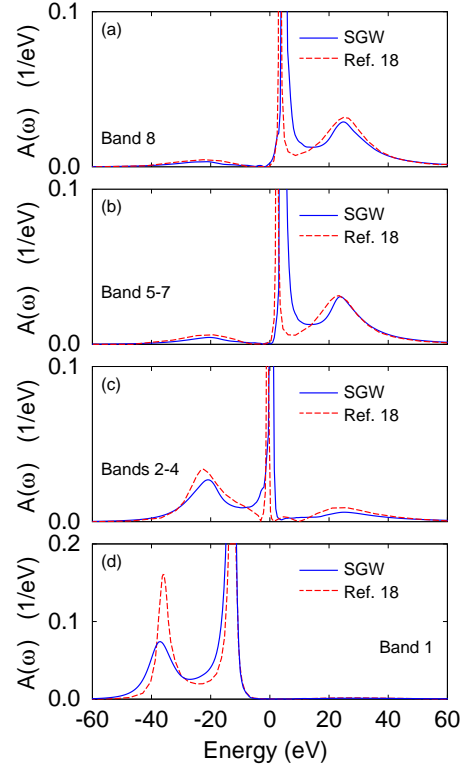}
\end    {center}
\caption{\label{fig6}
        (Color online)
        Expectation values of the quasiparticle spectral function $\langle n\k| A(\r,\rp;\w)|n\k\rangle$
        for the first 8 silicon eigenstates at $\k=0$.
        The solid blue lines are our S$GW$ results using the EPM, the dashed red lines are from the first-principles
        calculation of Ref.\ \onlinecite{blochl} using the LDA and the PAW method.
        The expectation values have been calculated within the diagonal approximation of Eq.\ (\ref{eq.specfun}).
        In each panel the sharp peak near the band extrema corresponds to a well-defined quasi-particle, while the two broad peaks
        corresponds to plasmarons, i.e.\ electrons or holes coupled to a cloud of real plasmons.\cite{hl}
        The suppression of one of the plasmaron peaks reflects the large imaginary parts of the self-energy in the
        corresponding panels of Fig.\ \ref{fig5}. We point out the different vertical scale in panel (d).
        }
\end    {figure}

\begin{table*}[th!]
\caption{\label{tab.2}
$GW$ quasi-particle excitation energies for the $\Gamma^\prime_{2c}$, $\Gamma_{15c}$, $\Gamma^\prime_{25v}$, $\Gamma_{1v}$
states of silicon, as well as for the conduction band bottom of silicon close to the $X_{1c}$ state.
We obtained the quasi-particle energies $E_{n\k}$ by solving the nonlinear equations
$E_{n\k} = \E_{n\k} + {\rm Re} [ \Delta\Sigma_{n\k}(E_{n\k})]$, with $\Delta \Sigma_{n\k} = \langle n\k| \Sigma - V_{\rm xc} |n\k\rangle$.
For completeness we also show the unperturbed eigenvalues calculated here within the EPM scheme,
and those of Ref.\ \onlinecite{blochl} within the DFT/LDA scheme for comparison.
It is interesting to observe that the value of the minimum band gap is not increased from
the EPM value of $0.82$ eV when applying the $GW$ correction. In absolute terms the $X_{1c}$ state shifts upwards by 0.41 eV upon
the $GW$ correction, but this shift is compensated by the concurrent upward shift of the $\Gamma^\prime_{25v}$
state by 0.44 eV.
This problem relates to the uncertainty on the XC potential within the EPM formalism, as discussed in Sec.\ \ref{sec.5d}.
\vspace{0.5cm}}
\begin{tabular}{l r@{.}l  r@{.}l  r@{.}l r@{.}l r@{.}l c}
\hline
\hline
   & \multicolumn{4}{c}{\hspace{-0.6cm} EPM/LDA eigenvalues}  &
\multicolumn{7}{c}{Quasiparticle energies} \\
   & \multicolumn{2}{c}{Present }  & \multicolumn{2}{c}{Ref.\ \onlinecite{blochl}} &
\multicolumn{2}{c}{\hspace{1cm}Present} & \multicolumn{2}{c}{\hspace{0.3cm} Ref.\ \onlinecite{hl86}} & \multicolumn{2}{c}{\hspace{0.3cm} Ref.\ \onlinecite{blochl}} &\hspace{1cm} Expt. \\
\hline
$\Gamma^\prime_{2c}$  &   3&89 & 3&23   & \hspace{1cm}  4&21 & \hspace{0.3cm}  4&08 &\hspace{0.3cm}   4&05 & \hspace{1cm}4.23$^{\rm a}$,4.1$^{\rm b}$\\
$\Gamma_{15c}$        &   3&42 & 2&54   & \hspace{1cm}  3&53 & \hspace{0.3cm}  3&35 &\hspace{0.3cm}   3&09 & \hspace{1cm}3.40$^{\rm a}$,3.05$^{\rm b}$\\
$\Gamma^\prime_{25v}$ &   0&00 & 0&00   & \hspace{1cm}  0&00 & \hspace{0.3cm}  0&00 &\hspace{0.3cm}   0&00 & \hspace{1cm}0\\
$\Gamma_{1v}$         & -12&62 & -11&97 & \hspace{1cm}-13&23 & \hspace{0.3cm}-12&04 &\hspace{0.3cm} -11&85 & \hspace{1cm}-12.5$\pm$0.6$^{\rm a}$\\
$X_{1c}$              &   0&82 & 0&55   & \hspace{1cm}  0&79 & \hspace{0.3cm}  1&29 &\hspace{0.3cm}   0&92 & \hspace{1cm} 1.17$^{\rm a}$\\
\hline
\hline
$^{\rm a}$ Ref.\ \onlinecite{r35}.
$^{\rm b}$ Ref.\ \onlinecite{r39}.
\end{tabular}
\end{table*}

Within the $GW$ method the values of the quasi-particle excitation energies are typically 
calculated by using first-order perturbation theory on the DFT eigenvalues.\cite{hl86}
The perturbation operator $\Delta \Sigma(\r,\rp;\w)$ corresponds to the difference between 
the $GW$ self-energy $\Sigma(\r,\rp;\w)$ and the DFT exchange and correlation potential $V^{\rm xc}(\r)$.
This approach is sensible because the complete quasi-particle equations
are similar to the ordinary Kohn-Sham equations if we replace the self-energy $\Sigma$
by the DFT XC potential $V^{\rm xc}$.\cite{hl} 

Within the EPM scheme the total potential $V^{\rm EPM}$ acting on the electrons is specified,\cite{cohen_berg}
but the electronic charge density is not connected to this potential
through a self-consistent procedure.\cite{appelbaum}
This limitation makes it difficult to identify an XC contribution within the empirical
pseudopotential. However, since the charge density obtained within the EPM method can be regarded 
as an approximation to the actual charge density,\cite{EPM-density}
it appears sensible to obtain the effective XC potential
as a functional of the EPM charge density
using the local density approximation.\cite{lda1,lda2}
This procedure is formally equivalent to assuming that the unscreened ionic pseudopotential
$V^{\rm ion}$ is given by $V^{\rm EPM} - V^{\rm Ha} - V^{\rm xc}$, where
the Hartree potential $V^{\rm Ha}$ and the XC potential $V^{\rm xc}$ are calculated using the EPM charge density.
{\it 
This uncertainty on the XC potential renders the calculation of the quasi-particle
excitation energies somewhat arbitrary, therefore the results presented in the following
should be regarded as qualitative and are presented only for the purpose of demonstrating 
a complete $GW$  calculation within our S$GW$ methodology.
}

Despite the above limitations, the expectation values of the
XC potential and of the exchange term of the $GW$ self-energy calculated here are surprisingly
close to those obtained in Ref.\ \onlinecite{hl86} using {\it ab-initio} pseudopotentials
at the DFT/LDA level. Indeed, for the valence band top 
$\Gamma_{25v}^\prime$ state and the conduction band bottom close to the $X_{1c}$ state
our calculated XC expectation values are -11.27 eV and -8.97 eV, respectively,
while Ref.\ \onlinecite{hl86} gives -11.80 eV and -9.61 eV for the corresponding states
at the DFT/LDA level. The agreement is even better when comparing the expectation values
of the bare exchange part of the $GW$ self-energy. In this case we find -12.43 eV and -5.07 eV
for the $\Gamma_{25v}^\prime$ state and the $X_{1c}$ state, respectively, to be compared
to the corresponding values of -12.54 eV and -5.28 eV of Ref.~\onlinecite{hl86}.
These results provide an {\it a posteriori} justification to our choice of calculating
the XC potential using the EPM charge density and the LDA functional.

Table \ref{tab.2} compares our
calculated quasi-particle excitation energies for electronic states at the $\Gamma$ 
point and for the conduction band edge of silicon with the results of Refs.\ \onlinecite{hl86,blochl}.
We find a good overall agreement between these different sets of calculations.
Taking into account that we are comparing our S$GW$ scheme
within the EPM implementation with more sophisticated {\it ab-initio} calculations,
such agreement is rather encouraging.

Figure \ref{fig6} shows the calculated quasi-particle spectral function 
  \begin{equation}\label{eq.specfun}
  \langle n\k| A|n\k\rangle = \frac{|{\rm Im} \Delta\Sigma_{n\k}|}
  {| \w-\E_{n\k} - {\rm Re}  \Delta\Sigma_{n\k} |^2 + | {\rm Im}  \Delta\Sigma_{n\k} |^2},
  \end{equation}
with $A(\r,\rp;\w)$ denoting the $GW$ spectral function and $\Delta \Sigma_{n\k} = \langle n\k| \Sigma - V_{\rm xc} |n\k\rangle$.
We note that Eq.\ (\ref{eq.specfun}) is obtained by assuming that the off-diagonal
matrix elements of the self-energy in the basis of the unperturbed
wavefunctions are negligible. A diagonal approximation is not always justified, nonetheless
we decided to adopt Eq.\ (\ref{eq.specfun}) to be consistent with Ref.\ \onlinecite{blochl}.
Figure \ref{fig6} shows good overall agreement between our calculations and the
LDA/PAW results of Ref.\ \onlinecite{blochl}. This comparison demonstrates
once again the validity of our methodology.

\section{Conclusions and outlook}\label{sec.conclusions}

The results presented in Sec.\ \ref{sec.results} demonstrate the feasibility
of the self-consistent Sternheimer approach to $GW$ calculations in the simple case of a prototype implementation
based on the empirical pseudopotential method. The extension of the present
methodology to {\it ab-initio} approaches based on norm-conserving\cite{ncpp}
or ultrasoft\cite{uspp} pseudopotentials should not present any difficulties
as the crucial issues in the calculation have already been addressed 
in this work. 
The main advantage of the present methodology consists of the definitive
elimination of the unoccupied electronic states from the calculations of
{\it both} the screened Coulomb interaction and the non-interacting Green's
function. Another appealing aspect is that our methodology constitutes a
generalization to frequency-dependent perturbations of 
density-functional perturbation theory,\cite{baroni.rmp} which is a well
established technique with a long history of successes.

As discussed in Sec.\ \ref{sec.scaling} the present approach is comparable
in performance to standard\cite{hl86} $GW$ techniques.
The question remains on whether it is possible to make significant
improvements over the methodology proposed here {\it without compromising on
the numerical accuracy}. The most time-consuming
step of the entire procedure is the application of the single-particle
Hamiltonian $\H$ to a search direction $\psi$ in the Hilbert space spanned
by the wavefunction basis set during the iterative solution 
of the linear systems in Eq.\ (\ref{eq.linsys.1}). In order to accelerate
this operation there are possibly three ways ahead: (i) the improvement of the
minimization algorithms adopted, (ii) the use of sparse representations of
the Hamiltonian, and (iii) the use of smaller basis sets. 

(i) The iterative solution of Eq.\ (\ref{eq.linsys.1}) is here performed
by first solving for $\Delta\psi_v^\pm$ at fixed $\Delta V$, and then 
by updating $\Delta V$ in the self-consistency cycle.
It should be possible, at least in principle, to combine these two
operations in a single minimization step. This could be achieved
for instance by using the variational formulation of density-functional
perturbation theory developed in Ref.\ \onlinecite{gonze-variational},
or by using simulated annealing techniques such as the Car-Parrinello 
method.\cite{carparrinello} Both of these approaches were developed
for Hermitian systems, hence appropriate generalizations
to non-Hermitian systems would be required to solve Eq.\ (\ref{eq.linsys.1}).
As mentioned in Sec.\ \ref{sec.conv} the total number of cBiCG iterations
required to solve Eq.\ (\ref{eq.linsys.1}) is typically 
$N_{\rm CG} N_{\rm SCF}\sim 100$, therefore the variational formulation
of DFPT or the Car-Parrinello minimization would be convenient
if they resulted in a significant reduction of the number of iterations
over this figure.

(ii) Another possibility for improving the procedure presented in this work
is to resort to a real-space representation of the kinetic energy operator
in the single-particle Hamiltonian. This can be achieved by
calculating the kinetic energy on a real-space mesh using finite-differences
methods.\cite{chelikowsky} The main advantage of this approach would be
to have $GW$ calculations scaling with the cube of the system size
$N_{\rm at}^3$ instead of $N_{\rm at}^3 {\rm log} N_{\rm at}$.
However, in actual calculations the numerical prefactor associated with this scaling
could be unfavorable. Indeed, the fast Fourier transforms used in a planewaves
representation to calculate the product of the potential and the wavefunctions
require $2\cdot 4 N_\r^{\rm den} {\rm log} N_\r^{\rm den}$ floating point operations,
while the cost of the real-space calculation of the Laplacian operator
to sixth order is $37 N_\r^{\rm den}$
(for a supercell with orthogonal axes).\cite{chelikowsky} 
While these estimates seem to speak in favor of the real-space method,
it is advisable to consider that the preconditioner used here\cite{tpa} for the 
cBiCG minimization cannot be simply adapted to real-space calculations.
In absence of effective preconditioners the planewave method
remains advantageous for essentially any relevant system size.

(iii) Another interesting option for improving our methodology is 
to drastically reduce the size of the basis set adopted. This could be achieved
for instance by adapting our implementation to electronic structure packages
exploiting local orbitals basis sets.\cite{siesta}
Interestingly the three possibilities here outlined are not mutually exclusive,
and probably an appropriate combination of all of these would eventually open
the way to the study of electronic excitations in
very large systems using the $GW$ method.

In summary, we propose a new methodology for performing $GW$ calculations
using the self-consistent Sternheimer equation (S$GW$). We show how
to calculate the screened Coulomb interaction and the non-interacting
Green's function without resorting to unoccupied electronic states.
We successfully demonstrate our method within a planewaves empirical
pseudopotential implementation and compare with previous studies for the
prototypical test case of silicon. In our method the standard generalzied plasmon-pole
approximation for the frequency dependence of the screened Coulomb interaction
has been replaced by a direct calculation along the imaginary frequency axis,
followed by an analytic continuation to the real axis.
In addition, we introduce the use of multishift linear system solvers for the simultaneous
calculation of multiple frequency responses at the cost of one single
iterative minimization.

It is our plan to adapt the present approach to deal with first-principles
pseudopotentials, and to explore the performance of our procedure in a local
orbital real-space implementation. We hope that this work will stimulate further effort
to develop improved methodologies for excited states calculations in large systems.

\begin{acknowledgments}
F.G. is grateful to Manish Jain for drawing Ref.\ \onlinecite{alavi} to his attention.
This work was supported by National Science Foundation Grant No. DMR07-05941 and by the Director, 
Office of Science, Office of Basic Energy Sciences, 
Materials Sciences and Engineering Division, 
U.S. Department of Energy under Contract No. DE-AC02-05CH11231. 
The research leading to these results has received funding from the European Research
Council under the European Community's Seventh Framework Programme (FP7/2007-2013) / ERC 
grant agreement no. 239578.
Computational resources were provided by the Oxford Supercomputing Centre.
\end{acknowledgments}

\appendix

\section{Preconditioned complex biconjugate gradient method}\label{app.cbcg}

In this work we solve the linear systems in Eq.\ (\ref{eq.linsys.1}) 
using the complex biorthogonal variant (cBiCG) of the conjugate gradient method (CG)
following Ref.\ \onlinecite{jacobs}. This method is an extension of the standard
conjugate gradients algorithm to the
case of general complex matrices.
As the cBiCG algorithm introduced in Ref.\ \onlinecite{jacobs} 
did not include preconditioning, in this appendix we describe 
the preconditioned version of the algorithm which we derived 
following Ref.~\onlinecite{golub}.
We are interested in solving the linear system
  \begin{equation}\label{eq.axeqb}
  Ax=b,
  \end{equation}
with A a complex linear operator (not necessarily Hermitian), $b$ a complex 
vector, and $x$ the unknown solution vector.
In the cBiCG algorithm of Ref.\ \onlinecite{jacobs} two sequences of residuals $r_n$ and
$\rt_n$ are generated in such a way that successive residuals 
are biorthogonal [i.e.\ $\<r_{n+1}|\rt_n\>=0$ and $\<\rt_{n+1}|r_n\>=0$].
Two sequences of search directions 
$p_n$ and $\pt_n$ are generated so that successive directions
are biconjugate [i.e.\ $\< A p_{n+1}|\pt_n \> =0$ and 
$\< A^\dagger \pt_{n+1}|p_n \> =0$].
The algorithm starts by setting the initial residuals to
$r_0 = b-Ax_0$ ($x_0$ being the initial guess for the solution vector $x$) 
and $\rt_0=r_0^\star$, and the initial search directions to $p_0=r_0$ 
and $\pt_0=p_0^\star$. Next the solution
vector, the search directions, and the residuals are updated at each
iteration as follows:
  \begin{eqnarray}
  \alpha_n & = & \<\rt_n|r_n\>/\<\pt_n|Ap_n\> \label{eq.cg1}  \\ 
  x_{n+1} & = & x_n + \alpha_n p_n \label{eq.cg2} \\ 
  r_{n+1} & = & r_n - \alpha_n Ap_n \label{eq.cg3} \\ 
  \rt_{n+1} & = & \rt_n - \alpha_n^\star A^\dagger \pt_n \label{eq.cg4}\\ 
  \beta_n & = & - \<A^\dagger\pt_n|r_{n+1}\>/\<\pt_n|Ap_n\> \label{eq.cg5}\\ 
  p_{n+1} & = & r_{n+1} + \beta_n p_n \label{eq.cg6}\\ 
  \pt_{n+1} & = & \rt_{n+1} + \beta_n^\star \label{eq.cg7} \pt_n. 
  \end{eqnarray}
The time-consuming step in this algorithm corresponds to the application of the operators
$A$ and $A^\dagger$ to the search directions $p_n$ and $\pt_n$. 
As there are two such operations per iteration, the computational complexity 
is twice that of the standard conjugate gradient algorithm.

The preconditioning of the linear operator can be achieved by left-multiplying the linear system 
in Eq.\ (\ref{eq.axeqb}) 
by $M^{-1}$: $M^{-1}Ax=M^{-1}b$. If we assume that the preconditioner $M$ can be 
written as $M=E^{\rm T}E$, then we can rewrite the system as follows:
  \begin{equation}
  E^{-1}AE^{-{\rm T}} E^{\rm T}x = E^{-1}b.
  \end{equation}
By defining $A^\prime=E^{-1}AE^{-{\rm T}}$, $x^\prime=E^{\rm T}x$, and $b^\prime=E^{-1}b$
we obtain the transformed system $A^\prime x^\prime=b^\prime$, for which the
standard cBiCG method applies.
While this procedure is formally correct, it is not convenient to explicitly
transform the linear operator.
It is convenient instead to rewrite the procedure in terms of
$A$, $b$, and $x$ by performing a few formal manipulations.
For this purpose we make the substitutions
$r^\prime = E^{-1}r$ and $p^\prime=E^{\rm T} p$. Some algebra leads straightforwardly
to the preconditioned version of the cBiCG algorithm:
  \begin{eqnarray}
  \alpha_n & = & \<\rt_n|M^{-1}r_n\>/\<\pt_n|Ap_n\>  \\
  x_{n+1} & = & x_n + \alpha_n p_n \\ 
  r_{n+1} & = & r_n - \alpha_n Ap_n \\ 
  \rt_{n+1} & = & \rt_n - \alpha_n^\star A^\dagger \pt_n \\ 
  \beta_n & = & - \<A^\dagger\pt_n|M^{-1}r_{n+1}\>/\<\pt_n|Ap_n\> \\ 
  p_{n+1} & = & M^{-1}r_{n+1} + \beta_n p_n \\ 
  \pt_{n+1} & = & M^{-1}\rt_{n+1} + \beta_n^\star \pt_n.
  \end{eqnarray}
The preconditioned  cBiCG algorithm needs to be intialized with
$r_0=b-Ax_0$, $p_0=M^{-1}r_0$, $\rt_0=r_0^\star$, and $\pt_0=p_0^\star$.
In this work we have used the Teter-Payne-Allan function as the preconditioner 
$M^{-1}$.\cite{tpa}

\section{Condition number of the linear system}\label{app.condition}

\subsubsection{Screened Coulomb interaction}

The iterative calculation of the screened Coulomb interaction through 
Eq.\ (\ref{eq.linsys.1}) at finite real
frequencies $\w$ can be considerably more time-consuming than in the static
($\w\!=\!0$) case. Simple tests indicate that 
the number of iterations required to achieve convergence increases with 
increasing frequency $\w$. This behavior suggests that the linear operator in
Eq.\ (\ref{eq.linsys.1}) becomes progressively more ill-conditioned 
as the frequency $\w$ increases.

In order to rationalize this observation, we here examine
the condition number of the linear operator $\H-\E_v\pm\w$ in Eq.\ (\ref{eq.linsys.1}).
The minimum number of iterations $N_{\rm min}$ required for the solution of
a linear system using the conjugate gradients algorithm is given by
  \begin{equation}\label{eq.cg}
  N_{\rm min} = \frac{1}{2}\sqrt{\kappa} \log(2/\varepsilon),
  \end{equation}
$\kappa$ being the condition number of the linear operator and $\varepsilon$ the
desired relative accuracy.\cite{golub} 
While the estimate Eq.~(\ref{eq.cg}) has
been derived for the original CG algorithm, we found empirically that it also 
provides a reasonable description of the convergence rate of the complex cBiCG version.
The condition number $\kappa$ of a linear operator can be calculated as the ratio 
of its largest to smallest eigenvalues.
For a given valence state 
$|v^\prime\>$ the linear operator $\H - \E_\vp + \alpha \P - \w$ 
in Eq.\ (\ref{eq.linsys.1b}) has the eigenvalues
$\E_v - \E_\vp + \alpha - \w$ and $\E_c - \E_\vp - \w$. 

Let consider first the simplest case where $\w=0$ and $\alpha>0$. In this case 
we find by inspection that the smallest eigenvalue is $\min(E_{\rm g}, |\alpha-W_{\rm occ}|)$, 
$E_{\rm g}$ being the fundamental energy gap and $W_{\rm occ}$ the valence bandwidth.
It is common practice to set $\alpha=2W_{\rm occ}$ to avoid null eigenvalues.
\cite{baroni.rmp} With this choice the smallest eigenvalue becomes $E_{\rm g}$.
On the other hand, the largest eigenvalue can be approximated by the
cutoff energy of the wavefunction basis set $E_{\rm cut}$.
In this case the condition number reads $\kappa = E_{\rm cut}/E_{\rm g}$.
As an example, if we are using a plane-waves basis with a kinetic enegry
cutoff of 30 Ry, we have an electron energy gap of 1 eV, 
and the our desired accuracy is $\varepsilon=10^{-10}$, then according to
Eq.\ (\ref{eq.cg}) the minimum number of iterations required to solve 
the linear system would be $N_{\rm min} = 240$. Empirical tests show that 
this estimate is quite accurate for the system considered in the present work.
In order to improve the convergence rate it is useful to employ preconditioning
techniques. We here adopt the Teter-Payne-Allan preconditioner\cite{tpa}
in order to ``compress'' the eigenvalue spectrum and thereby reduce the
condition number. An ideal preconditioner would make the linear operator
perfectly well conditioned ($\kappa=1$). In this case
the optimal  number of iterations (for a relative accuracy $\varepsilon=10^{-10}$) 
would be as small as $N_{\rm min,pc} = 12$. We have found empirically
that by using the Teter-Payne-Allan 
preconditioner the number of iterations
required to achieve convergence was in all cases in the range $N_{\rm TPA}=$15--40. 

We now consider the case of $\w<0$. Simple algebra shows that in this case
$\kappa = E_{\rm cut}/(E_{\rm g} + w)$
when $\alpha = 2W_{\rm occ}$. Hence in this case the larger the frequency $\w$,
the better conditioned the linear system. We checked this result
by explicit calculations.

The worst case in terms of condition number is found when $\w>0$. 
In fact, as soon as the frequency exceeds the optical excitation
threshold $\w>E_{\rm g}$, the linear operator acquires null eigenvalues 
corresponding to the resonance condition $\w = \E_c - \E_\vp$. 
In this latter case the condition
number $\kappa(\w)$ exhibits significant structure, reflecting
the joint density of states of the system. Even after preconditioning the system, 
the number of iterations required to achieve convergence can be as high as 
$N_{\rm min} = 500$, thus rendering this avenue unpractical.
The calculation of the screened Coulomb interaction for frequencies slightly
off the real axis $\w+i\eta$, with $\w>0$ and small $\eta$, leads to only a small
improvement of the convergence rate.
The difficulty of solving iteratively the linear system Eq.\ (\ref{eq.linsys.1})
for large positive frequencies is accompanied by the additional difficulty 
of adequately sampling the Brillouin zone to describe the singularities at $\w = \E_c - \E_\vp$.

Altogether these considerations suggest that an iterative solution of the
linear system along the real axis is not convenient from the computational
viewpoint. For this reason we decided to evaluate the screened Coulomb interaction 
along the imaginary axis and then to analytically continue the functions
to the real axis using {\it Pad\'e approximants}.\cite{pade1,pade2,blochl}
The motivation behind our choice becomes obvious after considering a simple
plasmon-pole model of the screened Coulomb interaction:\cite{hl86}
  \begin{equation}
  W(\w) = v + \frac{W_0 -v}{2} \Big[ \frac{\w_{\rm p}}{\w+\w_{\rm p}} - \frac{\w_{\rm p}}{\w-\w_{\rm p}} \Big],
  \end{equation}
where $\w_{\rm p}$ is the pole frequency and $W_0$ the static screened Coulomb interation.
Analytical continuation of this function to the imaginary axis yields
  \begin{equation} \label{eq.pp.im}
  W(\w=i\beta) = v + \frac{W_0 -v}{1+(\beta/\w_{\rm p})^2}.
  \end{equation}
Equation (\ref{eq.pp.im}) indicates that the screened Coulomb interaction along the imaginary axis contains the same
amount of information as the one on the real axis ($\w_{\rm p}$ and $W_0$), and at the same time
does not exhibit any singularities. In this case the condition number reads
(assuming no preconditioning and $\alpha=0$ for simplicity) 
$\kappa=[(E_{\rm g}^2+\beta^2)/(E_{\rm cut}^2+\beta^2)]^\frac{1}{2}$,
and tends to unity for large imaginary frequencies.
As a result, the worst case scenario for the solution 
of the linear system corresponds to the static case $\w=0$.

In summary, by solving iteratively the linear system along the imaginary
axis we circumvent the difficulties associated with the ill-conditioning
of the linear system in Eq.\ (\ref{eq.linsys.1}) occurring at real frequencies
and the necessity of dense Brillouin zone sampling.
The details of the analytic continuation are discussed in Appendix \ref{app.pade}.

\subsubsection{Green's function}

A similar analysis can be carried out for the calculation of the Green's
function using the method introduced in Sec.\ \ref{sec.green}.
It is straightforward to establish that in this case the condition number
of the system is given by $\kappa = E_{\rm cut}/\delta$.
As the infinitesimal $\delta$ is typically taken to be 0.1 eV, we are effectively
dealing with a situation analogous to a small-bandgap semiconductor.
The TPA preconditioner can be adopted to reduce the condition number
to $\kappa = E_{\rm kin}^{\rm VBM}/\delta$, where $E_{\rm kin}^{\rm VBM}$
is the expectation value of the kinetic energy of the highest occupied state
and is independent of the basis set cutoff. Numerical tests confirm that this 
is indeed a sensible and effective strategy.

\section{Analytic continuation using Pad\'e approximants}\label{app.pade}

In order to perform the analytic continuation of the screened Coulomb interaction
from the imaginary axis to the real axis, we employ diagonal 
Pad\'e approximants.\cite{pade1,pade2,blochl}
The Pad\'e approximant of order $N$ is the optimal rational approximation
to a target function $f(\w)$ known in $N$ distinct points 
$\w_n$,~$n=1,\cdots,N$. 
When $N$ is an odd integer the diagonal Pad\'e approximant reads
  \begin{equation}
  P_N(\w) = \frac{p_0+p_1\w+\cdots+p_{(N-1)/2}\w^{(N-1)/2}}
  {1+q_1\w+\cdots+q_{(N-1)/2}\w^{(N-1)/2}},
  \end{equation}
and its coefficients $p_0, p_1, \cdots, p_{(N-1)/2}$, $q_1, \cdots, q_{(N-1)/2}$ are determined by matching the approximant
to the target function in $N$ points $P_N(\w_n)=f(\w_n)$,~$n=1,\cdots,N$.
Both the coefficients and the Pad\'e approximant can be calculated
very efficiently using a simple recursive algorithm.\cite{pade2}
Some experimentation indicates that approximants of order $N\ge5$ are necessary
to reproduce a plasmon-pole spectral shape including a finite linewidth.
This observation can be rationalized by considering that a plasmon-pole
spectral function is completely defined by the values of the function at $\w=0$,
the location, strength, and width of the pole, and the asymptotic value at $\w=+i\infty$.
Some of this information is redundant and can be obtained by using sum rules.\cite{hl86}
The parity of the screened Coulomb interaction can also be exploited to minimize
the number of input frequencies. 
The advantage of the Pad\'e approximant is that a more refined description
of the frequency-dependent screened Coulomb interaction can simply be achieved
by calculating additional points along the imaginary axis. 

We also investigated the possibility of analytically continuing
the screened Coulomb interaction by using a multi-pole expansion
as suggested in Ref.\ \onlinecite{spacetime}. 
We tried one-, two-, and three-pole
expansions by determining the coefficients using the simplex 
method of Nelder and Mead.\cite{nelder-mead}
The single-pole approximation appears robust but the quality
of the real-axis continuation is poorer than what we obtained
by using Pad\'e approximants. Multi-pole approximations were found
to be unreliable because of their high sensitivity to the initial guesses 
for the coefficients.
Our experience therefore is that the multi-pole expansion is not an optimal 
choice for an automated procedure where the analytic continuation has to be
performed for every $\G$, $\Gp$, and $\q$ of the screened Coulomb 
interaction without manual intervention. 

\section{Simultaneous calculation of the susceptibility at multiple frequencies}\label{app.multishift}

The linear systems Eqs.\ (\ref{eq.linsys.1.nscf}) and (\ref{eq.green.3})
can be solved efficiently by using the ``multishift'' 
cBiCG method of Ref.~\onlinecite{frommer}.
Multishift methods exploit the knowledge gained during the iterative
solution of the {\it seed} system $Ax=b$ to determine the solutions 
of the {\it shifted} system $Ax+\w x=b$ with only a small computational overhead.
The rationale behind such method is that the seed system and the
shifted system share the same Krilov subspaces $\{b,Ab,A^2b,\cdots\}$,
therefore the residuals of the seed and of the shifted
systems can be taken to be collinear.\cite{frommer}

This multishift technique allows us to determine the
entire frequency-dependence of the dielectric matrix
by performing one single static calculation for each set of parameters $[\q$,$\G]$ in
Eq.\ (\ref{eq.linsys.4}).
For the seed system the algorithm is still given by Eqs.\ (\ref{eq.cg1})-(\ref{eq.cg7}).
For the shifted system we replace the calculation of the residuals
$r_{n,\w}$ and of the coefficients $\alpha_{n,\w}$, $\beta_{n,\w}$ corresponding
to the frequency $\w$ by the following relations:
  \begin{equation} \label{eq.shift.1}
  r_{n,\w}  =\frac{r_n}{\pi_{n,\w}};
  \hspace{0.2cm}
  \alpha_{n,\w}  =  \frac{\pi_{n,\w}}{\pi_{n+1,\w}}\alpha_n ;
  \hspace{0.2cm}
  \beta_{n,\w}  =  \Big(\frac{\pi_{n,\w}}{\pi_{n+1,\w}}\Big)^2\beta_n,
  \end{equation}
where the scaling factor $\pi_{n+1,\w}$ is calculated through the recurrence relation
  \begin{equation}\label{eq.shift.2}
  \pi_{n+1,\w} = (1+\w\alpha_n) \pi_{n,\w} + \frac{\alpha_n\beta_{n-1}}{\alpha_{n-1}}(\pi_{n,\w}-\pi_{n-1,\w}).
  \end{equation}
In order to obtain collinear residuals $r_n$ and $r_{n,\w}$ we need to
initialize the algorithm using $x_0=0$.

The use of Eqs.~(\ref{eq.shift.1}) and (\ref{eq.shift.2}) allows us to
skip the time-consuming operations involving the Hamiltonian in
Eqs.\ (\ref{eq.cg1}) and~(\ref{eq.cg5}).
This method is extremely convenient for determining the frequency-dependent
susceptibility for many frequencies at the cost of one single calculation.

We point out that this method still carries some drawbacks.
One limitation is that this method cannot be applied to the self-consistent
system of Eq.\ (\ref{eq.linsys.1}), because the know-term on the
right-hand side depends on the frequency $\w$ itself.
Therefore the use of the shifted cBiCG method is only possible
for {\it non self-consistent calculations of the dielectric matrix}
and requires explicit matrix inversions to determine the screened
Coulomb interaction.
This approach can be regarded as an improved version 
of the technique proposed in Ref.\ \onlinecite{reining-sternheimer}.

Another limitation is that the shifted cBiCG method
does not allow for the use of preconditioners. In fact
the preconditioned seed system $M^{-1}Ax=M^{-1}b$ and the preconditioned shifted system
$M^{-1}Ax+\w M^{-1}x=M^{-1}b$ do not share the same Krilov subspaces, 
hence the residuals cannot be taken to be collinear.\cite{simoncini} 
The practical consequence is that for systems with
large basis set energy cutoffs and small band gaps, the number of iterations
required to achieve convergence could be impractically large.


\begin{thebibliography}{99}
\bibitem{note.spinfactor}
The factor of 2 for the spin degeneracy is not included in Eq.\ (\ref{eq.green.1})
as only the states with the same spin couple when forming the 
matrix elements of the $GW$ self-energy.\cite{hl86}
\bibitem{note.integral}
Had we truncated the integral along the real axis as proposed for the Coulomb term, 
the $\Sigma^{\rm ex}$ term would have acquired a spurious frequency dependence associated with
the semicircle closing the complex integration path.
\end{thebibliography}
\end{document}